\documentclass[12pt,preprint]{aastex} 
\begin{document}      
\title{Detection, photometry and slitless radial velocities of
       535 planetary nebulae in the flattened elliptical galaxy 
       NGC 4697\footnote{Based on observations collected at the
       European Southern Observatory, Cerro Paranal, Chile; 
       ESO programme 63.I-0008}}
\author{R. H. M\'endez and A. Riffeser}
\affil{Munich University Observatory, Scheinerstr. 1,
       81679 Munich, Germany}
\email{mendez@usm.uni-muenchen.de} 
\author{R.-P. Kudritzki}              
\affil{Institute for Astronomy, University of Hawaii,
       2680 Woodlawn Drive, Honolulu, Hawaii 96822}
\author{M. Matthias}
\affil{Munich University Observatory, Scheinerstr. 1,
       81679 Munich, Germany}
\author{K. C. Freeman}
\affil{Mt. Stromlo and Siding Spring Observatories,
       Weston Creek P.O., ACT 2611, Australia}
\author{M. Arnaboldi and M. Capaccioli}
\affil{Osservatorio Astronomico di Capodimonte,
       V. Moiariello 16, Napoli 80131, Italy}
\and
\author{O. Gerhard}
\affil{Astronomisches Institut, Universit\"at Basel, CH-4102
       Binningen, Switzerland}               
                                   
\newpage
 
\begin{abstract}
We have detected 535 planetary nebulae (PNs) in the flattened
elliptical galaxy NGC 4697, using the classic on-band, off-band
filter technique with the Focal Reducer and Spectrograph (FORS) 
at the Cassegrain focus of the first 8-meter telescope unit of the 
ESO Very Large Telescope. From our photometry we have built the 
[O {\sc III}] $\lambda$5007 planetary nebula luminosity function 
(PNLF) of NGC 4697. It indicates a distance of 
$(10.5\pm1)$ Mpc to this
galaxy, in good agreement with the distance obtained from surface
brightness fluctuations and substantially smaller than a previous 
estimate of 24 Mpc used in earlier dynamical studies. The PNLF 
also provides an estimate of the specific PN formation rate: 
$(6\pm2)\times10^{-12}$ PNs per year per solar luminosity. 
Combining the information from on-band images 
with PN positions on dispersed, slitless grism images, we have 
obtained radial velocities for 531 of the 535 PNs. We describe 
the slitless velocity method, and the calibration procedures 
we have followed. The radial velocities have errors of about 
40 km s$^{-1}$ and provide kinematic information up to a distance
of almost three effective radii from the nucleus. Some rotation 
is detected in the outer regions, but the rotation curve of this 
galaxy appears to drop beyond one effective radius. Assuming
an isotropic velocity distribution, the velocity dispersion 
profile is consistent with no dark matter within three 
effective radii of the nucleus (however, some dark matter 
can be present if the velocity 
distribution is anisotropic). We obtain a blue mass-to-light 
ratio of 11. Earlier $M/L$ ratios for NGC 4697 were too small, 
because of the too large distance used for their derivation.

\end{abstract}
 
 
\section{Introduction}

Planetary nebulae (PNs) in early-type galaxies are extremely useful
as distance indicators, through their luminosity function (PNLF) 
(e.g. Jacoby, Ciardullo, \& Ford 1990; Jacoby 1997; M\'endez 1999;
Ferrarese et al. 2000), and also as test particles to study 
the dark matter distribution in the outskirts 
of those galaxies. A classic example of this second 
use is the work on 433 PNs in NGC 5128 by Hui et al. (1995). 
Similar attempts involving more distant galaxies observed
with 4m-class telescopes have resulted in very limited samples;
typically only 20 to 70 PN velocities could be measured (e.g. 
Ciardullo, Jacoby, \& Dejonghe 1993; Tremblay, Merritt, \& Williams 
1995; Arnaboldi et al. 1994, 1996, 1998). PN samples larger by at least 
one order of magnitude are needed for reliable dynamical analyses
(e.g. Merritt \& Saha 1993); the best results would probably be obtained
by combining a large PN sample with deep absorption-line spectroscopy,
using e.g. the method followed by Saglia et al. (2000).

The observational situation has recently improved.
Given a sufficiently luminous early-type galaxy, at a distance of 
10-15 Mpc, and working with 8m telescopes, now it is possible to 
detect 500 PNs and measure their radial velocities in a reasonable 
amount of time. The purpose of this paper is to describe these new 
capabilities and report on their first practical application.

FORS1, the first spectrograph to become available at the ESO VLT,
has some multi-object capabilities, supporting 19 slits. However 
this observing mode would not be suitable for a project involving  
hundreds of PNs. We decided to use FORS1 in a different way (slitless
radial velocities) to be explained in section 2.

As our first target we selected the flattened elliptical galaxy
NGC 4697, located in the Virgo Southern Extension. Concerning dark 
matter content, flattened ellipticals are almost unexplored. Most of
the information accumulated about dark matter in ellipticals comes 
from studies of hot, X-ray emitting gas (e.g. Loewenstein \& White 
1999) and from dynamical analyses of integrated absorption-line 
spectra of luminous, slowly rotating, nearly round ellipticals
(e.g. Gerhard et al. 2000). There is however an interesting
study of the flattened elliptical NGC 720 by Buote and Canizares
(1994, 1997), based on ROSAT and ASCA X-ray observations. These 
authors find evidence for a flattened dark matter halo around NGC 720.

According to Garcia (1993) NGC 4697 belongs to a group of 18 galaxies, 
but none of the other 17 is closer than 30 arc minutes to our target. 
X-ray observations with ROSAT (see 
Sansom, Hibbard, \& Schweizer 2000), suggest a rather small amount
of associated hot gas. More recent observations with Chandra
(Sarazin, Irwin, \& Bregman 2000) show that in fact most of the
detected X-ray emission is resolved into point sources, mostly
low-mass X-ray binaries. NGC 4697 must have lost most of its 
interstellar gas. Is this a hint that this galaxy does not have 
a substantial amount of dark matter? A dynamical study is, in this 
case, the only way to learn about the dark matter existence and 
distribution. Among the earlier dynamical studies of NGC 4697 we 
mention those by Binney, Davies, \& Illingworth (1990) and Dejonghe 
et al. (1996), both based on photometry and kinematic data from the 
central, high surface brightness region within one effective radius. 
The data were consistent with a constant mass-to-light ratio through 
the studied portion of the galaxy, but were not adequate to investigate
the dark matter distribution. In both papers a distance of 24 Mpc 
was assumed for NGC 4697. Since Tonry et al. (2000) have measured 
a surface brightness fluctuation (SBF) distance of 12 Mpc, we decided 
it would be interesting to determine a PNLF distance to NGC 4697.

In section 2 we describe the method of slitless radial velocity 
determinations. Section 3 describes the observations and reduction 
procedures. Section 4 deals with the on-band photometry. In section 5 
we build the PNLF and derive the PNLF distance to NGC 4697 and the
specific PN formation rate. In section 6 we describe 
the radial velocity results, which are analyzed in sections 
7 and 8. Section 9 gives a summary of conclusions and perspectives.

\section{Slitless radial velocities}


PNs in other galaxies can be detected by blinking two images: one taken 
through a narrow-band interference filter selected to transmit the 
redshifted [O {\sc III}] $\lambda$5007 nebular emission line (on-band image) 
and another one taken through a broader filter at some nearby wavelength, 
avoiding strong nebular emission lines (off-band image). The PNs 
are visible in the first image but not in the second one. The off-band
image must be deeper than the on-band image for the unambiguous detection
of emission-line sources.

Suppose that after the end of the on-band exposure a grism
is inserted in the light path. Nothing else is changed, 
the telescope is kept guiding, and we take a new exposure through 
both the grism and the on-band filter. Please refer to Fig. 1.
All stars in the field are shifted and become short 
segments of length defined by the grism dispersion and
by the width of the on-band filter transmission curve.  
The PNs (and any other emission-line point sources in the field) 
remain as point sources, also shifted from the position they had in 
the undispersed on-band image. The shift is a function of the  
wavelength of the nebular emission line and of position on the CCD
(the grism introduces anamorphic distortion). If we can measure and 
calibrate this shift, then it is possible to
calculate the emission-line wavelength for each detected PN, giving 
radial velocities for all PNs in the field with maximum efficiency,
irrespective of the number and distribution of PNs in the field. 
Since no slits are used, there are no light losses and it is not 
necessary to go through the complex selection and preparation 
procedures typical of multi-object slit (or fiber) spectroscopy. 
Of course the quality of the velocities will depend on the seeing. 
In ideal conditions, all the images needed to find the PNs and get 
the kinematic information can be collected in just one observing run.

The basic idea for slitless radial velocities is quite old.
A selected list of historical references can be found in Douglas
\& Taylor (1999). In recent times, slitless techniques for 
emission-line objects have been explored as alternatives 
to the classic multi-object technique. For example, work based 
on a Fabry-Perot interferometer (Tremblay, Merritt, \& Williams 1995)
and on counter-dispersed imaging (Douglas \& Taylor 1999; Douglas et al.
2000). Here we present the simplest procedure for slitless PN radial
velocities, applied for the first time with the light-collecting power
of an 8-meter telescope.


Let us describe the procedure for wavelength calibration. We need to 
determine the shift produced by the insertion of the grism as a 
function of wavelength and position 
on the CCD. Since we have 19 slits in FORS, we can use them for the 
calibration. We align the 19 slits vertically near the left edge of 
the CCD, and take an exposure illuminating the slits with the 
comparison lamp. This is the undispersed image. Then we insert the 
grism and take another exposure, the dispersed image. Now we move all 
the slits to the right, defining a new calibrating position, and repeat
the sequence of two exposures. This can be repeated until we have covered
the useful region of the CCD (since the grism produces a shift to the 
right, no calibration is possible for objects located near 
the right edge of the CCD). Fig. 2 shows the grid of calibration
positions (19$\times$10 slits) on the CCD, and Fig. 3 shows an 
example of undispersed and dispersed calibration images.


For all wavelength measurements, we assume that the direction of 
dispersion is exactly horizontal (which is not true, because there is 
a small amount of pincushion distortion; we will correct for this 
effect later) and measure the undispersed and dispersed $x$ positions 
of the slits at fixed $y$ values. With this information we can
find the relation between shift and wavelength for each of the 190 
calibration positions. The wavelength can be expressed as a polynomial 
of second order in the shift.

The procedure for PN wavelength calculation is the following: 
having measured the positions of a PN in the undispersed and dispersed 
images, select the four grid positions closest to the undispersed
PN coordinates. Calculate the wavelength for each of the four grid 
positions, using as input the shift between dispersed and undispersed
PN $x$ coordinates (acting as if the PN was located exactly at each of 
the four grid positions). From these four wavelengths, obtain the final 
wavelength by bilinear interpolation, using the correct undispersed PN
$x$, $y$ coordinates.


Our decision to assume a horizontal direction of dispersion is based 
on the lack
of any simple procedure that could help to determine the direction of 
dispersion for any point in the CCD. The price to pay for this 
simplification in the measurements is that we need a complementary
set of exposures to test for the presence of systematic errors in 
the radial velocities, arising from the pincushion distortion mentioned
above. We tested the accuracy of slitless radial velocities near the
center of the field by taking undispersed and dispersed images of 
He 2-118, a PN of small angular size with a sufficiently well-known 
radial velocity, $-164 \pm 9$ km s$^{-1}$ (Schneider et al. 1983).
Since it would be impractical to take many such exposures with the PN
in many different places across the CCD, we decided to make multi-object
through-slit (undispersed) images of NGC 7293, insert the grism, and take
the corresponding multi-object spectrograms. Six such sequences, with the
19 slits aligned at different x positions across the CCD, were enough 
for our purposes. We aligned the 19 FORS slits along a line shown in
Fig. 4. The radial velocity of NGC 7293 on this line is 
$-20 \pm 10$ km s$^{-1}$ (Taylor 1977). The 19$\times$6 slit positions
on the CCD are shown in Fig. 5. We found empirically that the radial
velocities must be corrected by a small term which is a function of the
distance from the measurement point to a reference diagonal line, also
shown in Fig. 5. Please refer to section 6, where we describe all the
slitless radial velocity results.

The accuracy of slitless radial velocities depends on being able to 
ensure that the distance between undispersed and dispersed images on 
the CCD is not modified by mechanical deformations in the spectrograph, 
or guiding problems. This is not relevant during the very short 
exposures needed for the calibrations, but it is a matter of concern 
when dealing with exposures longer than 30 minutes. We will discuss 
this potential source of error in section 6, showing that it is of 
minor importance in our case.

\section{VLT Observations and reductions}

The galaxy NGC 4697 has an effective radius of 95 arc seconds (Binney, 
Davies, \& Illingworth 1990). We decided to orient the x-axis of the
CCD in the direction of the major axis, and to expose two partly 
overlapping fields, NE and SW of the galaxy center, with the central 
region appearing in both fields, to provide some degree of redundancy.
We call the two fields E and W, for simplicity. They are shown in 
Figs. 6 and 7.

The observations were made by R.H.M. with the first Focal Reducer and 
low dispersion Spectrograph (FORS1) at the Cassegrain focus of 
Unit Telescope 1, Antu, of the ESO Very Large Telescope, Cerro Paranal,
Chile, on the nights of 1999 April 19/20 and 21/22, and 2000 May 29/30,
30/31 and May 31 / June 1. FORS1 with the standard collimator gives
a field of 6.8$\times$6.8 arc minutes on a 2080$\times$2048 CCD
(pixel size 24$\times$24 $\mu$m). The image scale is 0.2 arcsec/pixel.

Direct imaging was done through interference filters. The on-band and
off-band filters used for NGC 4697 have the following characteristics: 
effective central wavelengths, in observing conditions,
5028 and 5300 \AA ; peak transmissions 0.76 and
0.80; equivalent widths 48.5 and 215 \AA ; FWHMs 60 and 250 \AA.
The dispersed images were obtained with grism 600B. This grism gives
a dispersion of 50 \AA /mm, or 1.2 \AA /pixel, at 5000 \AA. 

Table 1 lists the most important CCD images obtained for this project. 
They can be divided into the following groups:

(1) off-band, on-band, (grism + on-band) images of NGC 4697, E field

(2) off-band, on-band, (grism + on-band) images of NGC 4697, W field

(3) on-band, (grism + on-band) multi-slit images and comparison lamp
    spectra through 10 different vertical arrangements of the 19 FORS 
    slitlets, for the wavelength calibration.

(4) on-band, (grism + on-band) images of the Galactic PN He 2-118
    (PN G 327.5+13.3) to test the quality of the slitless radial 
    velocities. In this case we used another on-band filter, with 
    effective central wavelength 4990 \AA \ and FWHM=60 \AA.

(5) on-band, (grism + on-band) multi-slit images and spectra of the 
    Galactic PN  NGC 7293 (PN G 036.1-57.1), in 6 different positions 
    across the CCD, for complementary calibration purposes. For this
    PN we also used the $\lambda$4990 narrow-band filter.

(6) on-band ($\lambda$5028) images of the spectrophotometric standard 
    G138-31 (Oke 1990) for the photometric calibration.

The 1999 April nights were partly clear but certainly not photometric.
The seeing was around 0.7 arc seconds. We were fortunate to get two 
photometric nights for our flux calibrations: the two first nights in 
May 2000. The typical seeing was 0.8 arc sec. The third and last night was
affected by sporadic cirrus clouds and poorer seeing (around 1 arc sec).

The basic CCD reductions (bias subtraction, flat-field correction
using twilight flats) were made using IRAF\footnote{IRAF is
distributed by the National Optical Astronomical Observatories,
operated by the Association of Universities for Research in
Astronomy, Inc., under contract to the National Science
Foundation of the United States} standard tasks.

Next, it was necessary to combine all undispersed on-band images, 
for E and W fields separately, to eliminate cosmic ray events and 
enable detection of faint PNs. Combined images were also made from 
the off-band images E and W and from the dispersed images E and W.

The image combinations required careful preparatory work. 
It is important to ensure that the distance in pixels between 
undispersed and dispersed PN images is not affected by incorrect 
registration of the images to be combined. For example, an error of 
0.3 pixel in the distance between undispersed and dispersed images 
produces an error of 20 km s$^{-1}$ in radial velocity, if the 
dispersion is 50 \AA /mm. The following procedure was followed:
first of all, for each field, E and W, one pair of (undispersed, 
dispersed) on-band individual images, of the best possible quality, 
taken consecutively at the telescope, was adopted as reference images. 
In this way any possible displacements produced by guiding problems or 
deformations in the spectrograph are reduced to a minimum. We had in 
total four reference images (see Table 1): one pair for the E field 
(30T00:51 and 30T01:17) and one pair for the W field (30T23:33 and
31T00:00). All the other images, including the off-band ones, were 
registered upon the corresponding reference image. 

For registration we used a program developed in Munich by C. A. G\"ossl
and A. Riffeser (2001, in preparation) as part of their 
``image reduction pipeline''. 
The registration program, given a sample of well-distributed reference 
stars in both the reference image and the image to be registered, allows 
rotations, linear shifts, and (very small) scale changes, in order to  
obtain a coordinate transformation that minimizes the residuals of all
reference stars, while preserving the point-spread-function (PSF) as much
as possible. The coordinates of the reference objects in all images
are obtained by a seven-parameter rotated Gaussian fit. For a star-like
PSF the fit is performed within a box of 21 $\times$ 21 pixels.

It turned out that the images taken in 1999 had to be 
slightly rotated for registration: this rotation angle was 0.2 degrees. 

The quality of the registrations showed interesting differences. The
registration of the undispersed E-field images was very good; using 30 
reference stars the residuals were never larger than 0.1 pixel. The 
registration of the undispersed
W-field images was not so good, particularly for the 1999 images requiring
rotation. Using 44 reference stars, most of the residuals were smaller than 
0.2 pixel, but some residuals for the 1999 images were as large as 0.3 pixel. 

The registration of dispersed images is more difficult. We had to model
the PSF within a larger box, 121 $\times$ 27 pixels, with a Gaussian
which is wider in the direction of dispersion.
The solutions for the E-field were based on 11 bright reference stars 
and produced residuals below 0.2 pixel.
The solutions for the W-field were based on 8 reference stars and
although typical residuals were below 0.2 pixel, in the case of the 1999
grism images some residuals were as large as 0.4 pixel.   

We produced the combined images, from the reference and 
registered images, using the IRAF task ``imcombine" with ``ccdclip" 
rejection algorithm. We made a test with the E and W combined
dispersed images: we compared the positions of bright PNs visible
in each reference dispersed image versus the PN positions in the
corresponding combined dispersed image. The comparison was satisfactory 
for the E-field, but we found that a correction was needed for the 
W-field. The reason for this correction will be explained in next 
paragraph. The correction (equivalent to about 50 km/s) was mapped 
using 30 bright PNs well distributed in the W field, and later on 
applied to the PN positions measured in the combined W dispersed image.
In this way we convinced ourselves that PN positions in the E and W
combined dispersed images were not affected by systematic errors larger 
than 0.3 pixel. Therefore this source of position errors contributes at 
most 20 km s$^{-1}$ to the total uncertainty in radial velocity. 
We will discuss the errors more extensively in section 6.

Why was a correction needed for the positions of PNs measured in the
combined dispersed W-field image? The transmission curves of 
narrow-band filters shift in wavelength as a function of ambient 
temperature, by about 0.2 Angstroms per degree Celsius. The positions 
of dispersed PN images are not affected, but the spectral segments of 
continuum sources are shifted if the temperature changes. 
From the information in the headers of dispersed images we know that
the dispersed images in the E-field were taken with temperatures 
between 11.4 and 12.8 degrees Celsius. This range of temperatures
is not enough to produce significant displacements. On the other hand
most of the dispersed images in the W-field were taken in 1999 with 
temperatures around 15.5 degrees Celsius, but the reference dispersed
image was taken in 2000 with a temperature of 12.4 degrees Celsius.
Since we used continuum sources for registration of the dispersed
images, there was a systematic difference in the PN positions, which 
we could detect and correct as explained above. 

We extract two conclusions from this experience: first, for the highest
accuracy in FORS slitless radial velocities it is better,
if possible, to avoid combining images taken in different observing 
runs, because very probably slight rotations will be needed for the 
registrations, leading to less satisfactory registrations; second, 
all grism exposures should be long enough to provide a sufficient 
number of bright PNs in the field for adequate control of the quality
of registration across the CCD. In future work we will try to register
dispersed images using directly the brightest PN images, avoiding in 
this way any dependence on ambient temperature variations.

For easier PN detection and photometry in the central parts of NGC 4697,
where the background is strongly varying across the field, we decided to 
produce images giving the differences between undispersed on-band and
off-band combined frames. In ideal conditions this image subtraction 
should produce a flat noise frame with the emission-line sources as the
only visible features. A critical requirement to achieve the desired 
result is perfect matching of the point spread functions (PSFs) of the 
two frames to be subtracted. For this purpose we applied a method for
``optimal image subtraction'' developed by Alard \& Lupton (1998),
as implemented in Munich by G\"ossl and Riffeser (2001, in preparation) 
as part of their ``image reduction pipeline''. Very briefly the idea is 
to split the images in many subimages (in our case a few hundreds), 
allowing for different convolution kernels in different subimages 
to match one image against the other. Figures 8 and 9 show sections of 
the resulting difference images in the E and W fields.

We cannot use this procedure for the dispersed
combined images because we do not have any off-band counterpart. 
Therefore in order to flatten the background and reduce the contamination 
of the fields by the stellar spectra, we had to use simpler methods. The
IRAF task ``fmedian'' was applied to the combined E and W dispersed images,
with a box of 17$\times$17 pixels, and the resulting medianed images were
subtracted from the unmedianed ones. Figures 10 and 11 show the result.
Although by no means perfect, the subtraction was sufficient for our PN
detections.

\section{PN detection and photometry}

The exposure times were adjusted to obtain similar counts in undispersed 
and dispersed combined images. This was achieved to within about 20\%.
Since our main scientific drive is to produce kinematic information from
the PNs, we require a detection in both the undispersed and dispersed 
combined images to consider a PN as confirmed. In addition we also require 
the object to be a point source and to be undetectable in the off-band 
image, to minimize confusion with emission-line sources in background 
galaxies. In fact several non-PN emission-line sources (brighter in the 
on-band but visible in the off-band, and/or extended) were detected; 
a list will be provided in a future paper, now in preparation.

We experimented with a variety of software for object detection,   
but finally decided to use the traditional blinking method, reasoning 
that anyway a lot of time would be needed to confirm by eye the reality 
of any automatic detections, particularly in the very difficult dispersed
images, because of the contamination by numerous stellar spectra, even
after subtracting the medianed image.

Each combined image to be blinked was subdivided into 100 subimages of 
220$\times$220 pixels, allowing some overlap between adjacent subimages.
The candidate PNs were found by blinking the on-band vs. off-band
subimages, and confirmed by blinking on-band vs. dispersed. In the
central regions of the galaxy the differences (on-band $-$ off-band) 
were blinked vs. dispersed images, and in this case the confirmation 
was provided by the overlap between the E and W fields. Near the center
it is difficult, in principle, to avoid contamination by emission-line 
sources in background galaxies. However the area affected is small and 
the degree of contamination is insignificant in view of the high surface 
density of PNs in that region. Besides, the surface brightness of the 
galaxy is high near the center, and any existing background sources  
are unlikely to be detected unless they are very bright.
The radial velocities provide an additional way of identifying sources 
not related to NGC 4697, as we will report in section 6.

The $x$, $y$ pixel coordinates of all the candidate PNs in the undispersed
and dispersed images were measured using the IRAF task ``phot'' with
centering algorithm ``centroid''. In Section 6 we will describe the 
procedure for radial velocity determinations; here we concentrate 
on the on-band photometry.

We will express our $\lambda$5007 fluxes in magnitudes $m$(5007),
using the definition introduced by Jacoby (1989):

$$   m(5007) = -2.5 \ {\rm log} I(5007) - 13.74    \eqno(1)   $$

\noindent For the flux calibration we adopted the standard star
G138-31 (Oke 1990). This star has a monochromatic flux at 5028 
\AA \ of 1.44$\times 10^{-15}$ erg cm$^{-2}$ s$^{-1}$ \AA$^{-1}$.
The flux measured through the on-band filter in erg cm$^{-2}$ s$^{-1}$
can be calculated knowing the equivalent width of the on-band filter; 
using equation (1) we find $m$(5007) = 19.15 for G138-31.

Since only part of the nights were photometric, and most PNs were
only measurable on the differences of combined images (on $-$ off), 
we had to proceed in several steps. First of all we used IRAF task 
``phot'' to make aperture photometry of G138-31. The FWHM of the 
G138-31 images was between 3 and 4.5 pixels. We adopted an aperture 
radius of 16 pixels; the sky annulus had an inner radius of 21 pixels 
and a width of 5 pixels. Exactly the same parameters were used to make
aperture photometry of a set of three moderately bright stars, visible 
in both fields E and W, on individual CCD frames 30T00:51, 30T23:33,
31T00:42, and 31T02:12 (see Table 1), taken on the May photometric 
nights 29/30 and 30/31. The three ``internal frame standards'' are 
between 2 and 3 mags fainter than G138-31, and they are located far 
from the center of NGC 4697, to avoid background problems. None of them
showed evidences of variability.

Having tied our spectrophotometric standard to the ``internal frame 
standards'', we could switch to strictly differential photometry. 
We made aperture photometry of our three internal standards on the
E and W on-band combined images, to determine a correction due to 
non-photometric conditions affecting some of the combined frames. 
On the same on-band combined images we subsequently made PSF-fitting 
daophot photometry (Stetson 1987; IRAF tasks ``phot'', ``psf'', 
``allstar'') of the 
three internal standards and of four bright PN candidates close to them 
(i.e. visible in both the E and W fields and far from the center of NGC 
4697). From the aperture photometry and psf-fitting photometry of the 
three internal standards we determined the aperture correction. Finally, 
we made psf-fitting photometry of all PN candidates on the difference 
images (on-band $-$ off-band), where of course no normal stars remain;
we used the four selected bright PNs to tie this photometry to that 
of the standards. From the good internal agreement through all these
steps we estimate internal errors in the photometry below 2\%.

A final correction is needed to obtain physical fluxes, which requires 
the on-band filter peak transmission (e.g. Jacoby, Quigley, \& Africano 
1987). We measured the filter transmission on 
multi-slit CCD frames illuminated with internal 
flat lamps and exposed (a) through the grism and (b) through 
the grism + on-band filter. From our measurements we estimate that any
uncertainties in the filter transmission correction cannot affect the 
final fluxes by more than 5\%, even taking into account that the filter 
transmission is to some minor extent a function of position on the CCD. 

After measuring the observed wavelengths of the detected emission lines
(please refer to section 6) we verified that all PNs have redshifted 
wavelengths between 5022 and 5034 \AA. None of them is shifted away 
from the flat peak of the on-band filter transmission curve, 
irrespective of PN position on the CCD, and therefore we have not
made any correction to the photometry as a function of redshift.

As a further test we have used the redundancy provided by the overlap 
between the E and W fields: plotting magnitude differences 
between the two measurements (E and W) of PN candidates as a function 
of difference in distance from the center of the CCD, we have found a 
scatter diagram without any evidence of correlation.

Figure 12 shows the E on-band magnitudes $m$(5007) as a function of 
the W on-band magnitudes for 211 objects measured in both fields. 
From the dispersion in Fig. 12 we estimate RMS errors of 0.1 and 0.2
mag for $m$(5007) brighter and fainter than 26.5, respectively. For each 
of these 211 PNs we have adopted the average of the two $m$(5007)
measurements. 

\section{The PNLF, distance, and PN formation rate}

Knowing the apparent magnitudes $m$(5007) we can build the PN luminosity 
function. Since we are dealing with a non-uniform background, we need 
to produce a statistically complete sample, because the detectability 
of a PN varies with the background surface brightness. We operate in 
the following way: first we find the value of $m$(5007)
at which the raw, uncorrected PNLF starts to show a decreasing number
of PNs per bin, as we move towards fainter magnitudes. This indicates
the onset of severe incompleteness (see e.g. Fig. 7 in Jacoby et al. 
1989, or Fig. 2 in Jacoby et al. 1990). In our case, 
this ``incompleteness limit" is at $m(5007) = 27.6$. 
All PNs fainter than 27.6 are subsequently ignored for PNLF purposes. 
If the resulting sample is statistically complete, then the surface 
density of PNs brighter than 27.6 anywhere in the galaxy must be 
proportional, in a first approximation, to 
the surface brightness of the galaxy at that location. We attribute 
any departure from this proportionality (i.e. lack of PNs) to 
incompleteness. In doing so we disregard a small color effect: 
the surface density of PNs normally decreases with increasing
$B-V$ (see e.g. the discussion in section 8 of Hui et al. 1993). Since
NGC 4697 shows a small color gradient (e.g. Table IV in Peletier et al.
1990), the PN surface density should decrease slightly towards the 
center. By ignoring this color effect we are in fact slightly 
overestimating the incompleteness.

We test for completeness by counting the number of PNs brighter than 
27.6 detected in squares of a suitable size (we adopted 220$\times$220 
pixels) and plotting this number as a function of the background 
counts per pixel for each square. We show this plot in Fig. 13. 
As expected, initially the number of detected PNs grows proportional 
to the background counts, until the relation ceases to be linear 
somewhere above 1000 counts per pixel. Therefore to ensure 
completeness we build the PNLF using only those PNs brighter 
than 27.6 that are sufficiently distant from the galaxy center 
to have background counts below 1000 per pixel. The zone of exclusion
is an ellipse at the center of NGC 4697, with (minor, major) semiaxes 
of (200, 300) pixels (the image scale is 0.2 arcsec/pixel). 
We are left with 328 PNs, which is more than 
enough for a good PNLF distance determination.

The statistically complete PNLF is plotted in Fig. 14, which shows the
absolute magnitudes $M$(5007) derived using an extinction correction of
0.105 mag (the $E(B-V)$=0.03 is taken from Tonry et al. 2000, who used a
recent study of Galactic extinction by Schlegel, Finkbeiner, \& Davis
1998), and a distance modulus $m-M$ = 30.1. The observed PNLF is fitted 
with a simulated PNLF like the one used by M\'endez \& Soffner (1997)
to fit the observed PNLF of M 31. 
The simulated PNLFs plotted in Fig. 14 are binned, 
like the observed one, into 0.2 mag intervals, and have the 
following characteristics: maximum final mass 0.63 solar masses, 
$\mu_{max}$=1, and sample sizes between 2500 and 4900 PNs (see 
M\'endez \& Soffner 1997; the ``sample size'' is the {\it total\/} 
number of PNs, detected or not, that exist in the surveyed area).
The fit in Fig. 14 permits an unambiguous determination of distance and 
sample size because the observed PNLF shows an evident change of slope.

Attempts to fit the observed PNLF at a distance modulus of 29.9 or 
30.3 fail clearly. We conclude that our derived distance modulus of
30.1 has an internal error of 0.1 mag. 

Now we estimate the total error budget. Following the careful
discussion in Jacoby et al. (1990), we distinguish between
``possible systematic'' and random errors. The net systematic
error is exactly the same as in Jacoby et al. (1990), i.e. 0.13
mag, including the possible error in the distance to M 31, in the
modeling of the PNLF and in the foreground extinction. The random
contributions in our case are: 0.1 mag from the fit to the PNLF,
as mentioned above; 0.05 mag from the photometric zero point; 
and 0.05 mag from the filter calibration. If we combine all 
these errors quadratically we conclude that the total error bar 
for the distance modulus must be $\pm$0.18 mag. 

We find good agreement, within the uncertainties, with the SBF 
distance modulus 30.35 $\pm$ 0.2 reported by Tonry et al. (2000). 
Our distance modulus is equivalent to a distance of 10.5 Mpc.
Note that the small difference between the PNLF and SBF distances 
is in the same sense as for other galaxies more distant than 10 Mpc; 
above that limit SBF distances are systematically larger than PNLF 
distances (see M\'endez 1999).

It is also possible to get a PNLF distance using the analytical
representation (exponential + cutoff) proposed by Ciardullo et al.
(1989). This is shown in Fig. 15. The shape of the analytical PNLF 
is somewhat different, but the resulting distance modulus is the same.

After estimating the sample size in Fig. 14 we can calculate the 
specific PN formation rate $\dot \xi$, 
in PNs per year per solar luminosity, using the following expression:
 
$$ n_{\rm PN} = \dot \xi \ L_T \ t_{\rm PN}    \eqno(2)   $$
 
\noindent where $n_{\rm PN}$ is the sample size, $L_T$ is the total
bolometric luminosity of the sampled population, expressed in solar
luminosities, and $t_{\rm PN}$ is the lifetime of a PN, for which we 
have adopted 30\,000 years in the PNLF simulations.

We need to estimate the sampled luminosity in NGC 4697. From
$B_T$=10.14, $B-V$=0.89 (de Vaucouleurs et al. 1991), $A_B$=0.13
(Tonry et al. 2000) and a bolometric correction of $-0.78$ mag,
we obtain an extinction-corrected apparent bolometric magnitude 
8.34. Using the distance modulus 30.1 and a solar 
$M_{\rm bol}$ = 4.72, the total bolometric 
luminosity of NGC 4697 is $3.9\times10^{10}$ solar luminosities.
Since we omitted the central ellipse to obtain our statistically 
complete PN sample, and estimating from our images that the central 
ellipse contributes 50\% of the total luminosity, we conclude that
the sampled luminosity is $L_T = 1.95\times10^{10}$ solar 
luminosities. Adopting $ n_{\rm PN} = 3500$, we get 
$\dot \xi = (6\pm2)\times10^{-12}$. 

Transforming our value of $\dot \xi$ into a specific PN 
density $\alpha_{2.5}$ in PNs per solar luminosity, as used 
e.g. by Jacoby et al. (1990) and Ciardullo (1995), we find  
$\alpha_{2.5}=15\times10^{-9}$. Thus NGC 4697 fits 
quite well into the empirical relation between $\alpha_{2.5}$ 
and absolute $B$ magnitude (see Fig. 2 of Ciardullo 1995).

\section{Radial velocities: results and discussion}

In section 2 we have explained the procedure to measure slitless 
radial velocities, and here we present the results. In what follows
we will refer to heliocentric radial velocities determined with the 
slitless method simply as ``velocities''. We consider first the 
velocities of NGC 7293 measured at the 114 positions shown in Fig.5. 
All velocities were correct to within 50 km s$^{-1}$, but closer 
inspection showed a trend as a function of position on the CCD. 
This trend cannot be attributed to the velocity field in NGC 7293, and 
must be due to the pincushion distortion produced by the FORS optics.
We discovered empirically that this behavior could be modeled as
a cubic parabola if we plotted the velocities as a function of the 
distance from the undispersed position to the diagonal reference line 
shown in Fig. 5. These distances are defined to be positive above 
the reference line, and negative below; see Fig. 16. We introduced 
an empirical correction designed to give the expected velocity 
($-20$ km s$^{-1}$) 
irrespective of position on the CCD. The result of applying the 
correction is shown in Fig. 17. Our calibrations 
give velocities with errors below 20 km s$^{-1}$. We
verified this using the test exposures of the PN He 2-118: the 
velocities obtained from observations in two different nights are
$-175$ and $-184$ km s$^{-1}$, in good agreement with the catalog
velocity ($-164$ km s$^{-1}$). 

Now concerning NGC 4697:
if we add quadratically the calibration errors and the errors from
image registration (Section 3) we get errors of about 30 km s$^{-1}$.
However there is another source of errors: spectrograph deformations
and guiding problems during long exposures. The only way of testing
the impact of this kind of errors is to compare velocities obtained
from different pairs of (undispersed, dispersed) long exposures.  
For this purpose the redundancy between the E and W fields becomes
very useful. In Fig. 18 we compare the velocities of 165 PNs 
measured in both fields. The standard deviation in Fig. 18 is
36 km s$^{-1}$. We conclude that spectrograph deformations and 
guiding errors contribute only marginally, if at all, to the total
uncertainty in the velocities, which we estimate to be  
40 km s$^{-1}$. We have adopted the average of both velocity 
measurements for the 165 PNs in Fig. 18.

Most of the velocities of PNs in NGC 4697 are distributed between 
900 and 1600 km s$^{-1}$. We found three emission-line point sources
that gave very discrepant velocities if interpreted as $\lambda$5007
emitters: 2100, 2050, and 1800 km s$^{-1}$. We rejected these three
objects as PNs, assuming that they are high-redshift, background 
emission-line galaxies. Of course our PNLF plots do not include 
them either. These three sources will be described in a future
paper, now in preparation, which will bring a full catalogue of the
535 PNs in NGC 4697 and a list of non-PN sources. It is conceivable 
that there are a few additional contaminants which we could not
identify as non-PNs; but their possible existence will not affect 
the conclusions we present in what follows.

Fig. 19 shows the $x$, $y$ coordinates of the 535 PNs relative to the 
center of light of NGC 4697. The position of this center can be 
measured with errors below 1 pixel. The PN distribution has the 
shape of a rectangle:
there must be more PNs beyond the limits of our survey. However,
to detect and study them will be quite time consuming, because 
for low PN surface densities there will be more contamination with
background sources, and the only way to confirm any on-band point 
source as a PN will be to take a deep spectrum and detect the pair 
of [O {\sc III}] emission lines $\lambda\lambda$4959, 5007.

Fig. 20 shows the velocities of 531 PNs as a function of the $x$ 
coordinate relative to the center in pixels. Four PNs were too far
to the right of the W field to permit a velocity measurement. Figure
21 shows the same 531 velocities as a function of the $y$ coordinate.
The average velocity of the 531 PNs is 1270 km s$^{-1}$, in good
agreement with the most accurate velocities quoted in the literature 
for NGC 4697, namely 1236, 1250 and 1307 km s$^{-1}$ in de Vaucouleurs 
et al. (1991), da Costa et al. (1998), and Trager et al. (2000), 
respectively. Our velocity for NGC 4697 has an uncertainty of about
15 km s$^{-1}$, if we take into account a velocity dispersion of
the order of 150 km s$^{-1}$ (see section 8 below); the number of PNs 
we measured; and the possible systematic error of $\pm10$ km s$^{-1}$ 
in our velocities from the calibration procedure using NGC 7293.

\section{Rotation}

Since we have the CCD x-axis oriented in the direction of the major 
axis of NGC 4697, we first investigate the rotation simply by 
defining 8 subsamples along the x-axis and calculating the average 
velocity for each subsample (the subsamples are not equally spaced
in $x$ because we want to ensure a minimum number of PNs per 
subsample). The result is shown in Fig. 22, compared
with rotation data derived from long-slit spectroscopy along the major 
axis in the inner region of the galaxy by Binney et al. (1990). Their
effective radius $R_{\rm e}$ = 95 arc seconds is 
equivalent to 475 pixels in our CCD. They
did not find any rotation along the minor axis (that is to say around 
the major axis). We have subdivided the PN data into four subsamples 
along the minor axis, confirming that there is no evidence of rotation 
in that direction; the average velocities from the 4 subsamples are, 
in order of increasing $y$ values, 1244, 1280, 1278, and 1260 km s$^{-1}$. 

On the other hand, as expected, the PNs in Fig. 22 indicate some 
rotation along the major axis. The PNs appear to follow the sense of 
rotation indicated by the absorption-line studies: there are noticeable 
deviations in the average PN velocities at 0.5 $R_{\rm e}$, where 
absorption lines indicate maximum rotation velocity. The
signature of rotation from the PNs is diluted by the extension of 
our 8 subsamples in the $y$ direction: Binney et al. (1990) measured
slower rotation along lines parallel to, but 10 and 20 arc sec away 
from the major axis. In order to illustrate this dilution effect,
Fig. 23 shows the result of defining PN subsamples restricted
to a smaller range of $y$ values. The new 
PN subsamples cover a slot 20 arc sec wide along the major axis. 
A narrower slot would give too small PN subsamples. Now the 
average PN velocities follow the absorption line data much more 
closely, except near the center, where some dilution of rotation 
is still evident. But Fig. 23 indicates that the PNs behave 
kinematically like the stellar population.

Consider now the outer regions: the extreme points in Figs. 22 
and 23 give a difference in velocity of about 90 km s$^{-1}$, 
a rather small quantity if we compare with the 220 km s$^{-1}$ 
derived inside from the absorption-line data. The rotation curve 
of this galaxy appears to drop beyond one effective radius. 

\section{Line-of-sight velocity dispersion and $M/L$ ratio}

In order to study the run of the line-of-sight velocity dispersion as a 
function of the angular distance from the center, we defined a central
zone of radius 70 arc seconds and 2 intermediate rings, with boundaries 
at 140 and 200 arc seconds. Each of these zones was subdivided into SW 
(receding) and NE (approaching), to minimize the effect of rotation on 
the dispersions (since we know from previous work that in this galaxy 
dispersion is more important than rotation, we do not expect any
remaining residual rotation to affect our conclusions concerning 
dispersion). To the 6 central zones we added two outer zones, again 
SW and NE, with PNs having $x$ coordinates larger than 200 arc seconds 
relative to the 
center. The numbers of PNs within each zone are listed in the caption 
to Fig. 24, where we show the resulting line-of-sight velocity 
dispersions, which have been corrected to compensate for the effect 
of measurement errors of 40 km s$^{-1}$. We also show the velocity 
dispersions derived from absorption-line spectra along the major axis 
by Binney et al. (1990). In the central region the integrated light 
data and PN data are in excellent agreement; the dispersion from PNs
is slightly (but not significantly) larger.

We derive an estimate of the total mass and mass distribution of NGC 4697 
using an analytical model proposed by Hernquist (1990). This model is
spherical, nonrotating and isotropic, and we use it just as a first
approximation to the dynamical problem. The run of the line-of-sight
velocity dispersion predicted from this model is shown in Fig. 24.
The fit was obtained adopting a total mass of 1.9$\times 10^{11}$
solar masses and $R_{\rm e}$=95 arc seconds, which is equivalent to 
4823 pc for the distance modulus $m-M$=30.1. We tried to fit both the
absorption-line data and the PN data; a fit using only the PN data
would have required a total mass of 2.0$\times 10^{11}$ solar masses.

Now we can estimate the $M/L$ ratio in blue light: knowing the 
extinction-corrected $B_T$=10.0, the distance modulus 30.1 and the solar
$B$ absolute magnitude 5.48, we obtain for NGC 4697 a blue luminosity of
1.7$\times 10^{10}$ solar luminosities, which gives $(M/L)_{\rm B}$=11.
In earlier studies (e.g. Dejonghe et al. 1996), based on more adequate
2-integral and 3-integral models, the adoption of a too large distance
(24 Mpc) led to a too small $(M/L)_{\rm B} \sim 5$. We have verified that 
the Hernquist model gives, for a distance of 24 Mpc, the same $M/L$ ratio
obtained by Dejonghe et al. We have also followed a simple dynamical
analysis similar to that used by Hui et al. (1995) in their study of
NGC 5128, finding again a total mass of about 2$\times 10^{11}$
solar masses for NGC 4697.

A blue $M/L$ ratio near 10 is consistent with the predictions from 
stellar population models like those of Maraston (1998, and private
communication) for ages of the order of 10 Gyr if we adopt a Salpeter 
IMF and a metallicity slightly higher than solar (according to recent 
studies by Trager et al. (2000) and Thomas \& Maraston (in preparation) 
the [Z/H] ratio of NGC 4697 is near 0.1).  

Since the Hernquist model assumes a constant $M/L$ ratio, the good fit 
in Fig. 24 indicates that we have found no evidence of dark matter 
within 3 $R_{\rm e}$ of the galaxy's nucleus. This result is consistent 
with the apparent drop of rotation velocity in the outer parts, 
mentioned in the previous section, and with the absence of  
hot gas, mentioned in the introduction.

Note however that the Hernquist model assumes an isotropic velocity 
distribution. If there is anisotropy (e.g. a preponderance of radial
orbits outside) then some dark matter can be present.
To make a more careful study of the dynamics of NGC 4697 
is beyond the scope of this paper. In the near future we expect to 
make VLT+FORS long-slit spectroscopy of NGC 4697 at different
positions and position angles, extending the absorption-line studies 
to 2 $R_{\rm e}$ if possible. Combining this additional information 
with the PN data we can then make a more detailed study based on 
3-integral models adapted to a flattened system. That will be the 
subject of a future paper.

\section{Recapitulation and perspectives}

We have detected 535 PNs in NGC 4697 and measured their brightnesses and
radial velocities. We have built the [O {\sc III}] $\lambda$5007 PNLF and
used it to estimate the distance to NGC 4697, $(10.5\pm1)$ Mpc, in 
good agreement with the surface brightness fluctuation distance, and 
its specific PN formation rate, $(6\pm2)\times10^{-12}$ PNs per year 
per solar luminosity. 

We have completed the first large-scale experience with a method of 
slitless radial velocity determination for emission-line sources,
based on direct on-band plus dispersed grism+on-band exposures. 
The method is efficient and easy to calibrate, producing velocities 
for all detected emission-line sources in the field, irrespective of 
their number and distribution, 
with errors of the order of 40 km s$^{-1}$ in the 
specific case of VLT+FORS. The errors may become smaller if all 
observations are made within a single observing run, because one of 
the main sources of uncertainty is the error in registration when 
slight rotations of the dispersed frames are needed. 

We have provided PN kinematic information covering the full galaxy up 
to a distance of almost three effective radii from the nucleus. Some 
rotation is detected in the outer regions, but the rotation curve of 
this galaxy appears to drop beyond one effective radius. Assuming an 
isotropic velocity distribution, the velocity dispersion profile is 
consistent with no dark matter within three effective radii of the 
nucleus (however, some dark matter can be present if the velocity 
distribution is anisotropic). A more detailed dynamical study of 
NGC 4697 is postponed until we secure additional integrated light 
data. We obtain a blue mass-to-light ratio of 11. Earlier $M/L$ 
ratios for NGC 4697 were too small, because of the too large 
distance used for their derivation.

The work we have presented opens several interesting possibilities
for future projects. Dynamical studies based on PNs can now be extended 
to many field and cluster elliptical galaxies in the distance range from 
10 to at least 20 Mpc, without all the trouble associated with planning 
and executing multi-object spectroscopy for hundreds of sources. Our
empirical knowledge of dark matter distributions in ellipticals will
certainly increase. In addition, given enough telescope time, it will
be possible to make abundance determinations for many of the brightest 
detected PNs; for example, we would need 10 to 15 hours of total 
exposure time with FORS to measure abundances in the 10 brightest 
PNs of NGC 4697. It would seem that we are getting close to being able
to explore a PN abundance approach to metallicity gradients as well as
to the still unsolved problem of age-metallicity degeneracy in 
elliptical galaxies.

\section{Acknowledgments}

We thank the anonymous referee for careful reading and useful comments.

\newpage
 
\notetoeditor{Fig. 1 is intended for the full width of the
page (two columns). Figs. 8 and 9 are intended to fill one 
even-numbered page, and Figs. 10 and 11 the following odd-numbered 
page, in such a way that the reader can see all four of them 
simultaneously. All the other figures (2-7, 12-24) fit in one column.}
 
\figcaption[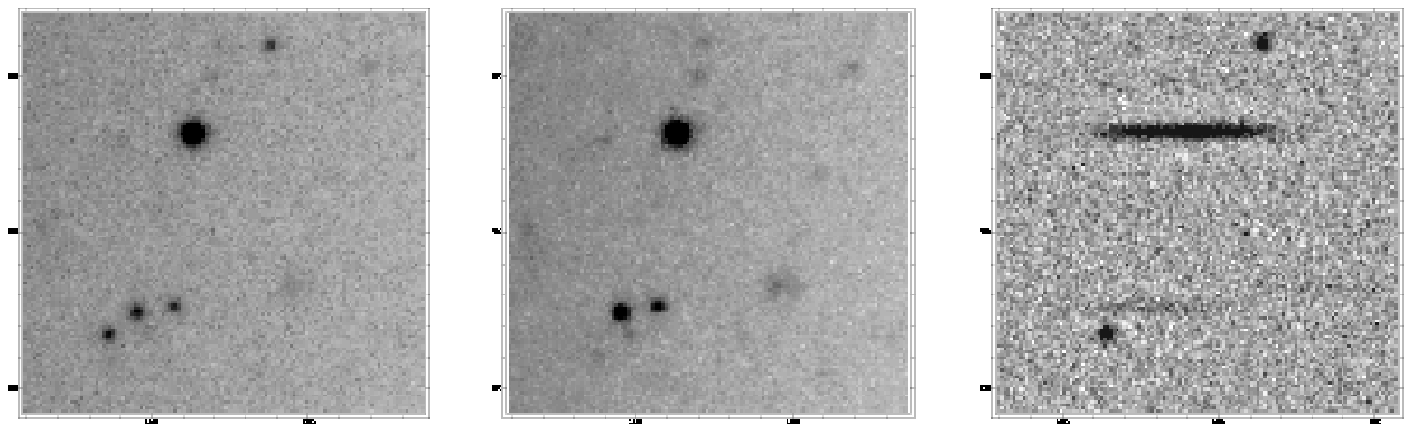]{On-band (left), off-band (center), and 
dispersed (grism + on-band) (right) subimages taken from the W field 
combined FORS1 exposures of NGC 4697 (see section 3). The area shown is 
130$\times$130 pixels, and is located 750 pixels west from the center 
of NGC 4697. Two PN candidates are visible in the on-band, absent in 
the off-band, and remain as point sources in the dispersed image, but 
shifted some 330 pixels to the right along the direction of dispersion. 
This shift, once calibrated, gives the wavelength of the detected 
emission line.
\label{fig1}}

\figcaption[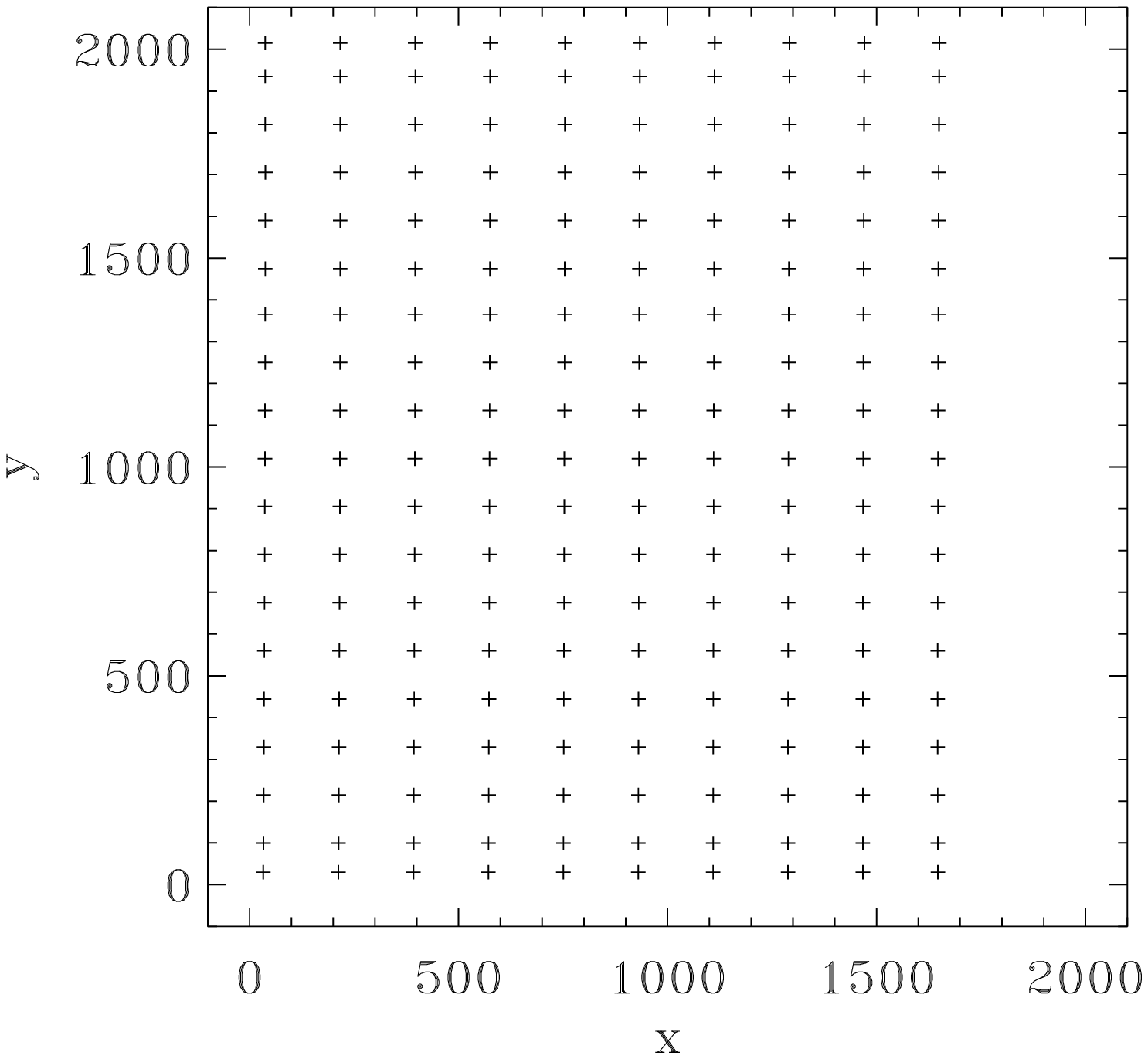]{The plus signs indicate the $x$, $y$ coordinates, 
in pixels, of the 190 undispersed slit positions used for the 
wavelength calibration. For slits at $x>$1650 pixels, 
the comparison spectrum falls partly outside
of the CCD and no reliable wavelength calibration is possible.
\label{fig2}}

\figcaption[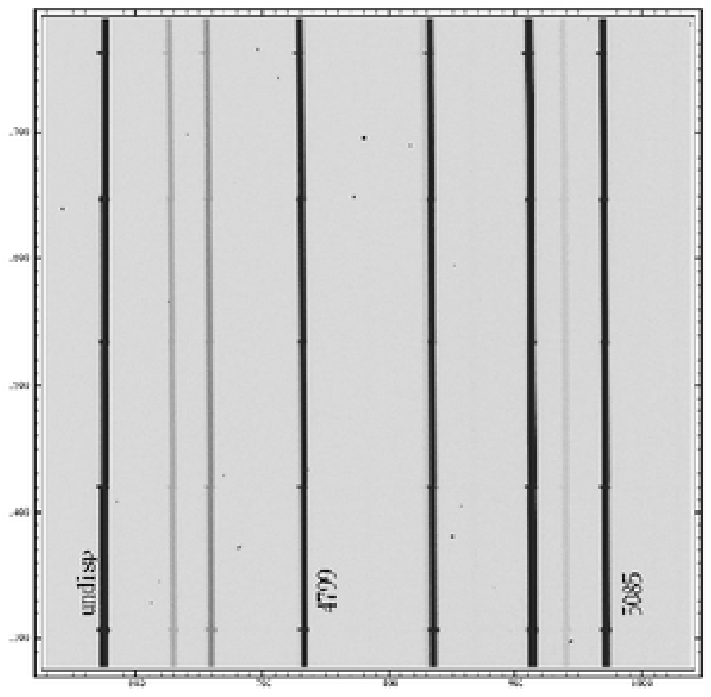]{To save space we have added images 30T12:48
(undispersed) and 30T12:50 (dispersed), which correspond to calibration 
position 4 (see Table 1). Here we show a small portion of this added 
image. The separation between slits is very small, and when the slits
are vertically aligned they look like a continuous line.
At the extreme left we see the undispersed images of 4 of the 19 
vertically aligned slits; the dispersed images to the right correspond 
to Cd and He comparison lines at 4678, 4713, 4799, 4921, 5015 and 5085 
\AA. These 6 lines were used for wavelength calibration at all the grid 
positions shown in Fig. 2. The weak comparison line at 5047 \AA \ was 
not used.
\label{fig3}}

\figcaption[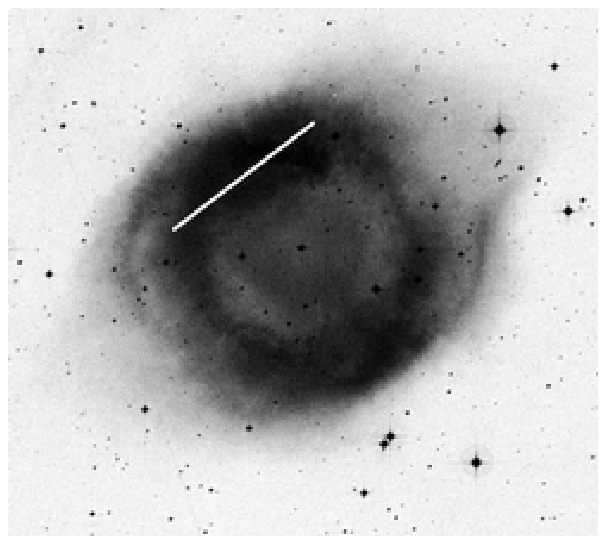]{On this image of NGC 7293, taken from the Digitized 
Sky Survey, we have indicated with a white line the approximate position
of our 19 FORS slits. The telescope was pointed so that this part of the 
nebula fell on 6 different $x$ coordinates across the CCD (see Fig. 5).
The radial velocity of NGC 7293 on the white line 
is $-20\pm10$ km s$^{-1}$.
\label{fig4}}

\figcaption[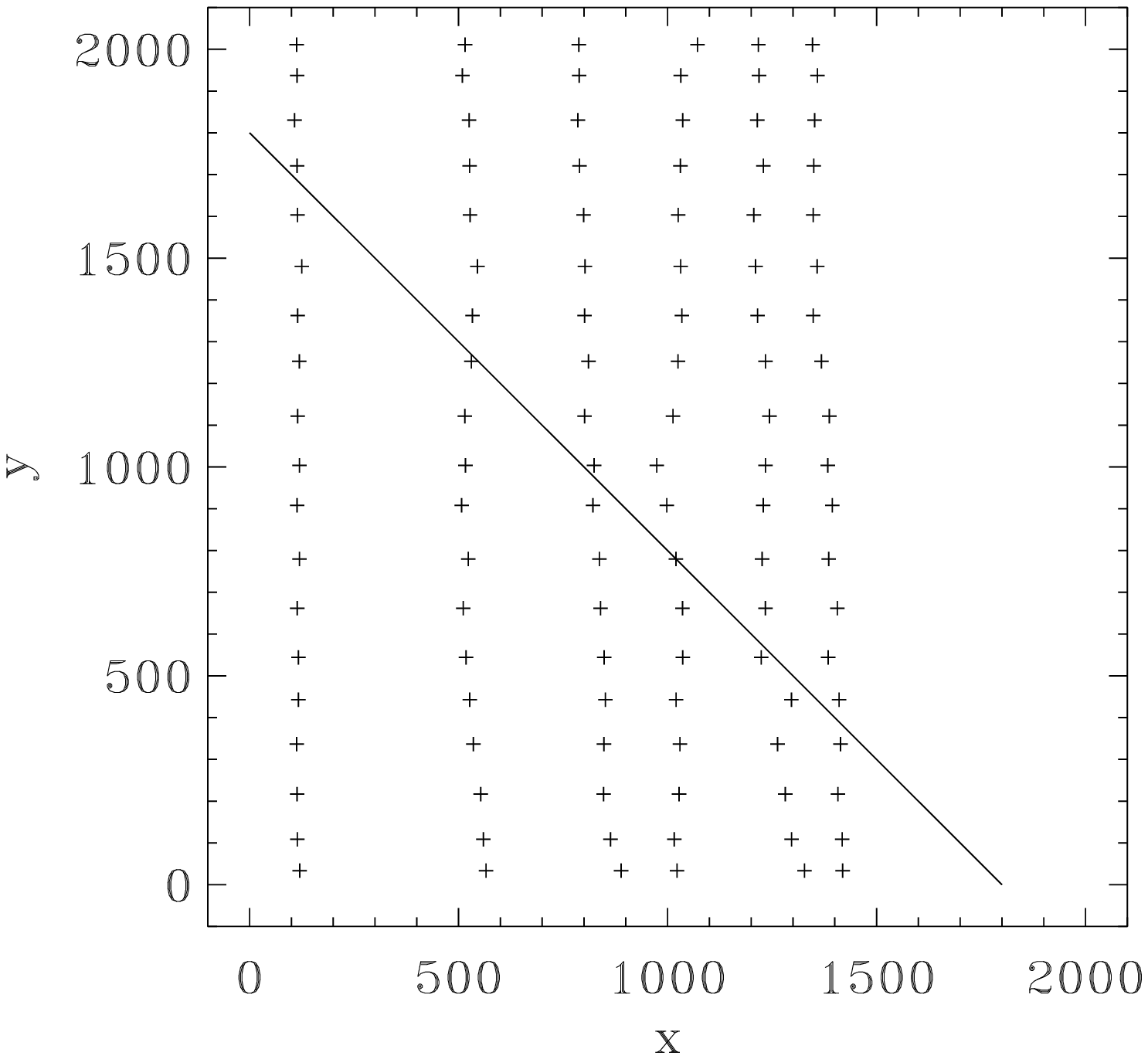]{The plus signs indicate the $x$, $y$ coordinates, 
in pixels, of the 114 (19$\times$6) undispersed slit positions used 
in NGC 7293. In this case no effort was made to produce a perfect 
vertical alignment. The descending diagonal is a reference line
used to calculate the small distortion corrections (see Section 6).
\label{fig5}}

\figcaption[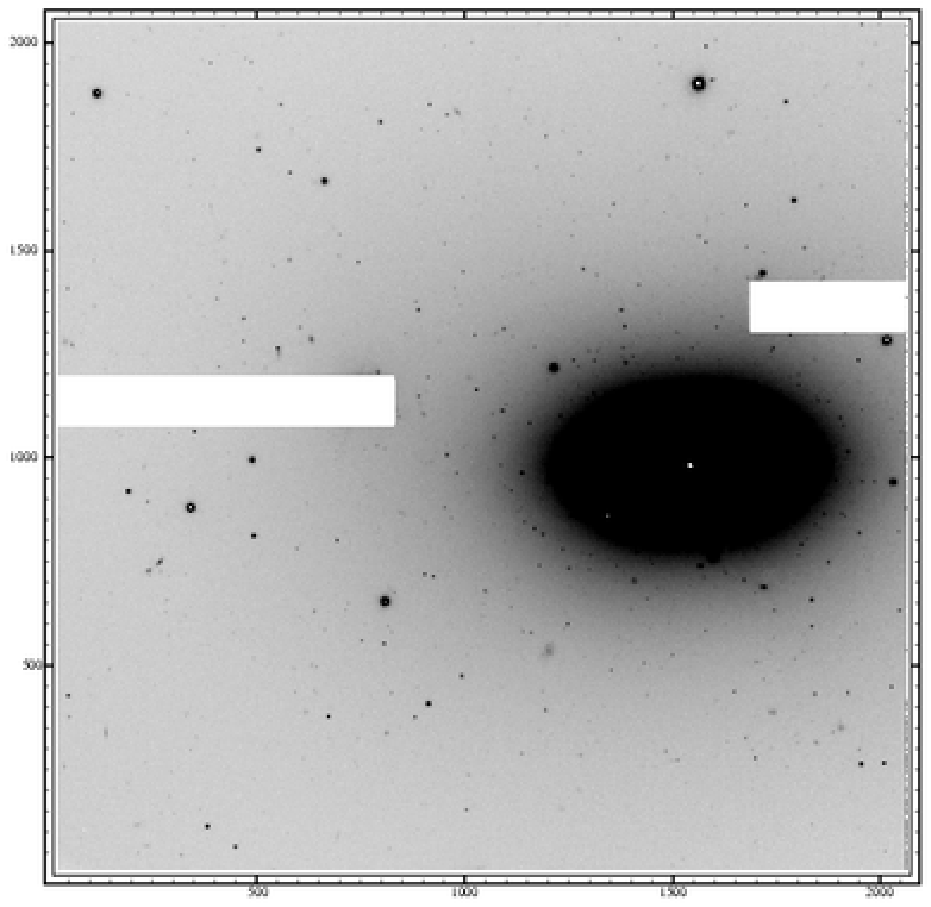]{NGC 4697, on-band combined image, Field E. 
Two of the blades used in FORS to define slits for multi-object 
spectroscopy were used as ``fingers'' to cover very bright stars 
in the field. The size of this field is 6.8 $\times$ 6.8 arc
minutes. The x-axis of the CCD is oriented in the direction of the 
major axis of the galaxy, at a position angle of 66 degrees (from 
N through E). Therefore, using compass names, in this figure NNW is  
up, and ENE is to the left.
\label{fig6}}

\figcaption[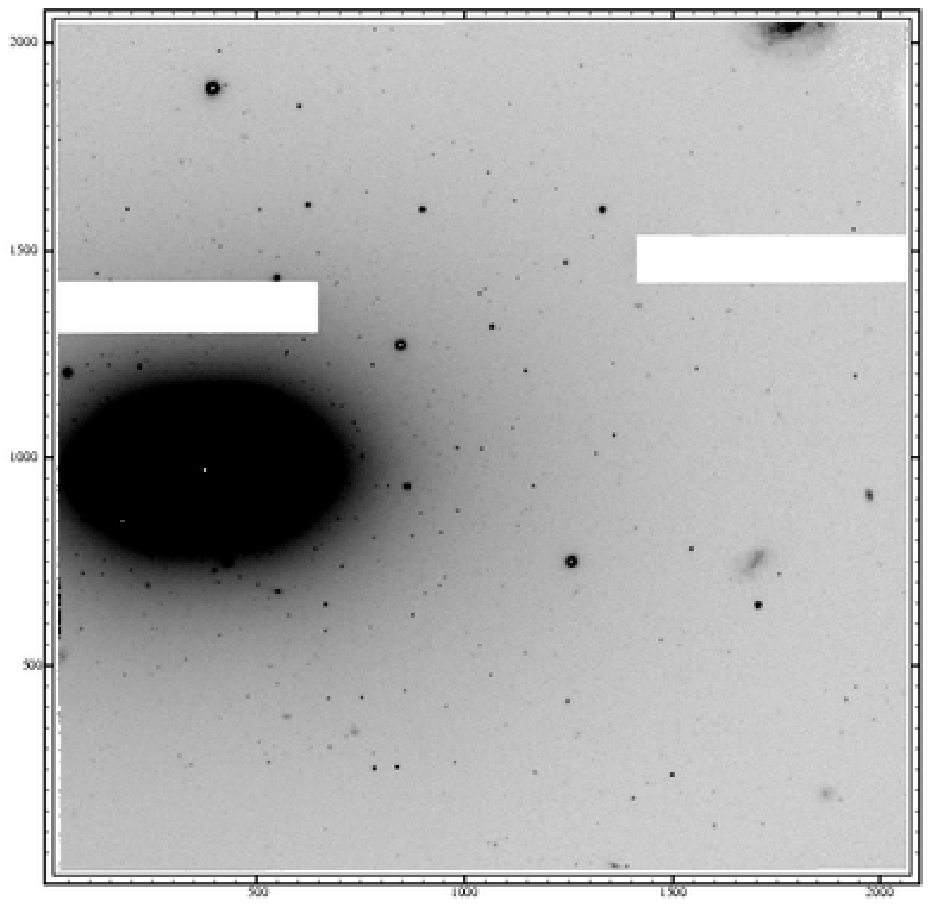]{NGC 4697, on-band combined image, Field W. 
Two ``fingers'' were used here as well. Size and orientation of the 
field as in Fig. 6.
\label{fig7}}

\figcaption[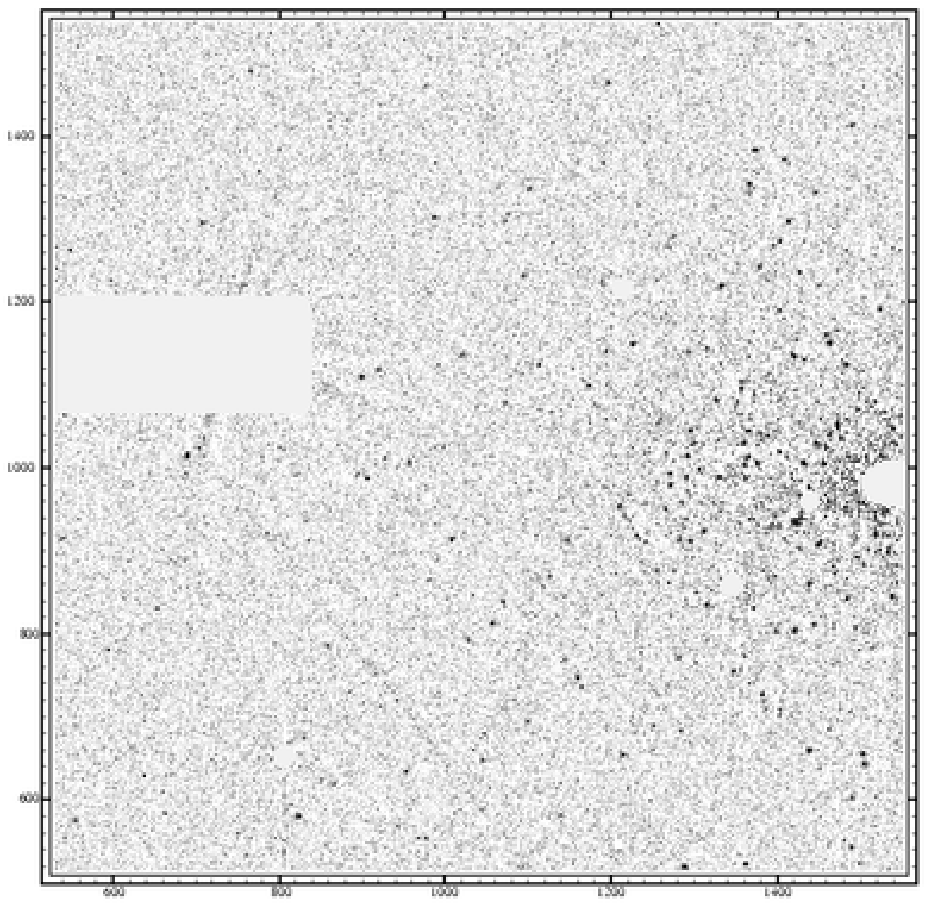]{NGC 4697, on-band $-$ off-band difference image, 
part of Field E. Some parts of the image near saturation level were 
replaced by zeros; the ellipse marks the center of the galaxy. The PNs 
appear as black dots.  
\label{fig8}}

\figcaption[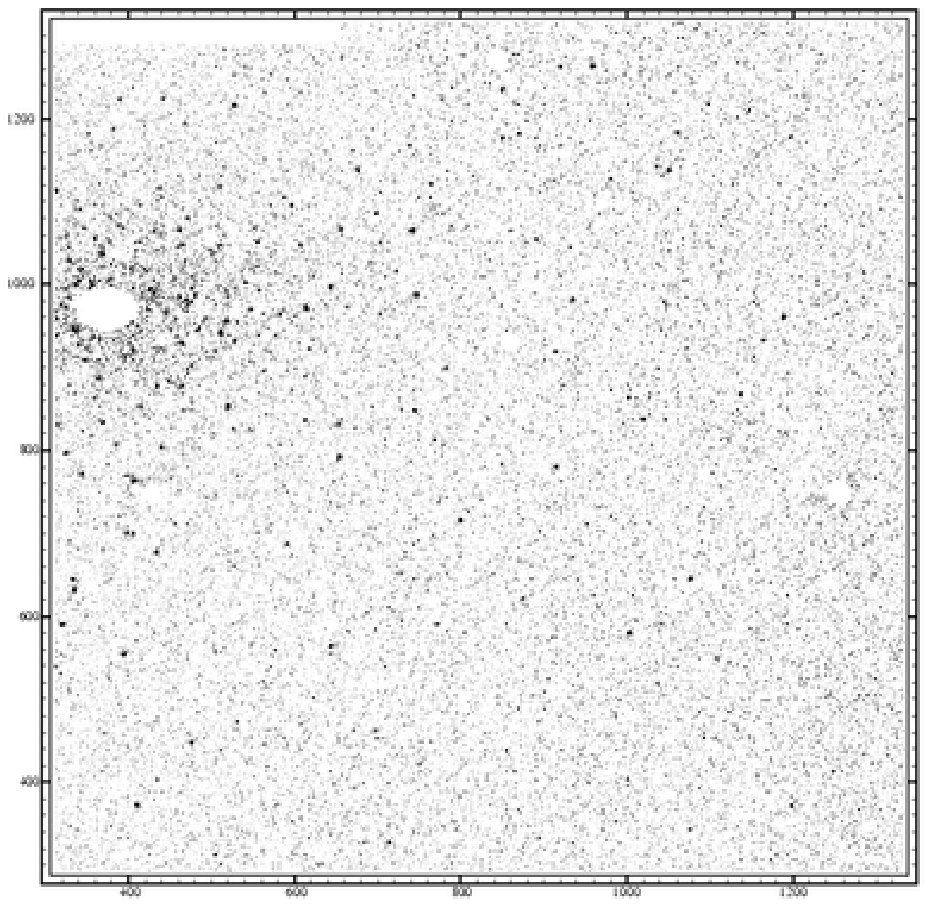]{NGC 4697, on-band $-$ off-band difference image, 
part of Field W. Same description as in Fig. 8.
\label{fig9}}

\figcaption[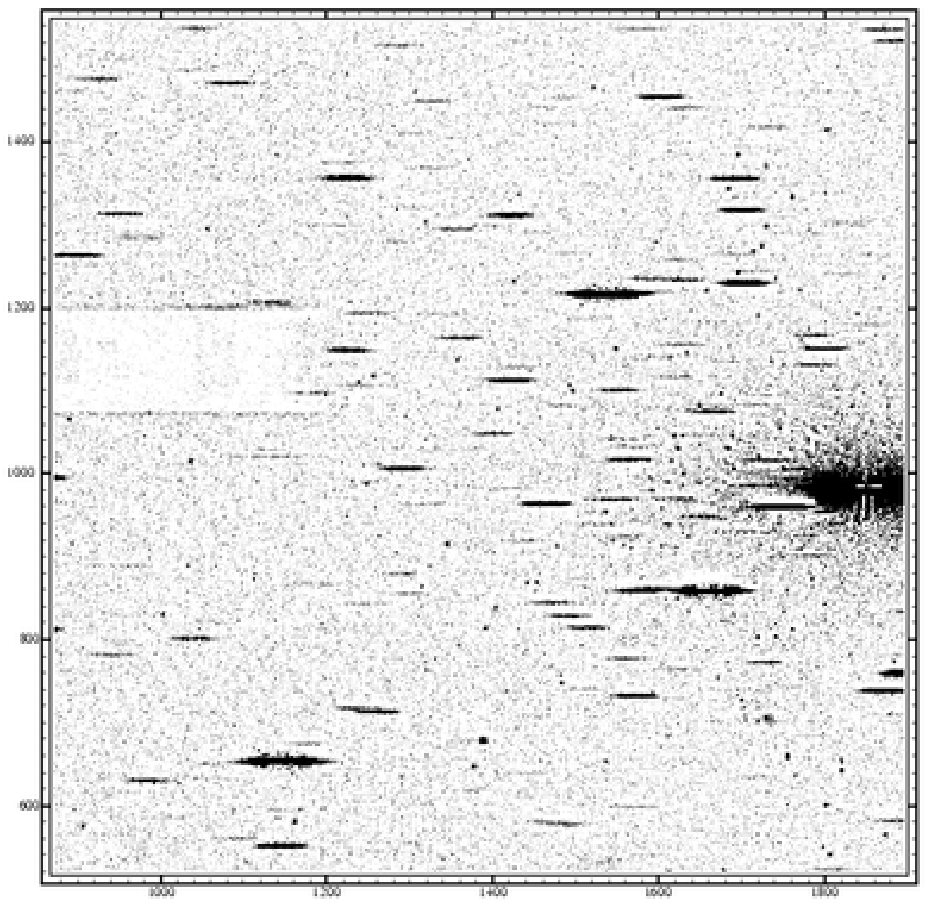]{NGC 4697, unmedianed $-$ medianed grism image,
Field E (same part of galaxy as in Fig. 8). 
\label{fig10}}

\figcaption[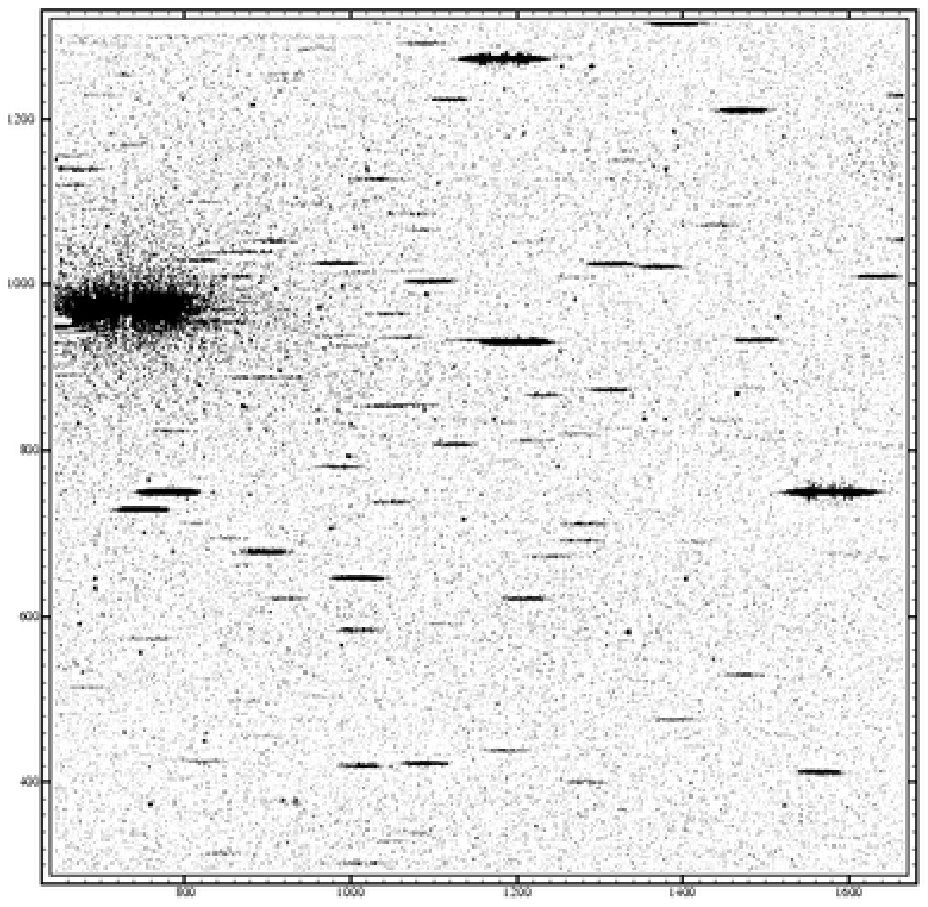]{NGC 4697, unmedianed $-$ medianed grism image,
Field W (same part of galaxy as in Fig. 9).  
\label{fig11}}

\figcaption[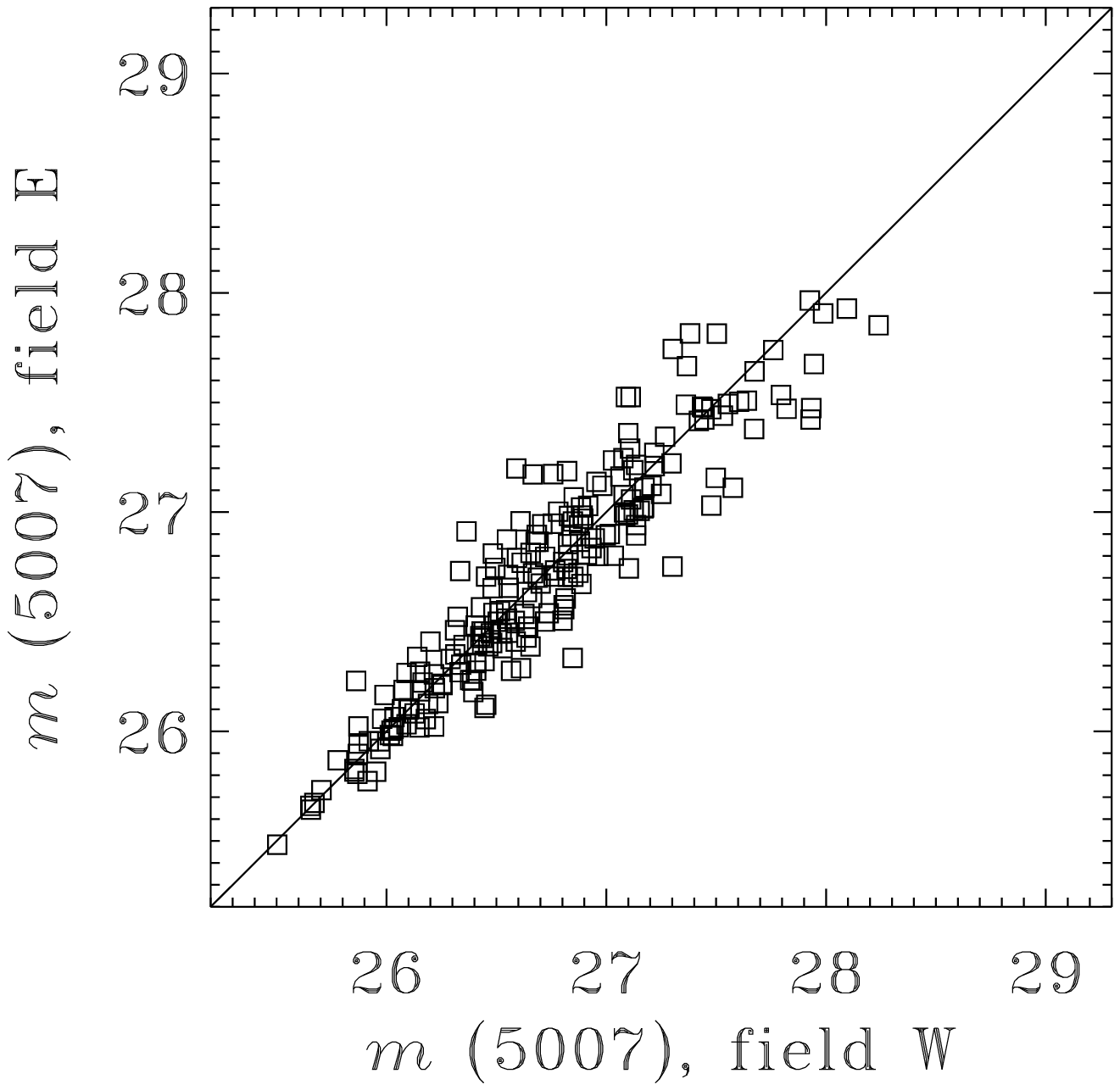]{The mags $m$(5007) of 211 PNs were measured in 
both fields, E and W. Here we compare both measurements. From the
dispersion we estimate RMS errors of 0.1 and 0.2 mag 
for $m$(5007) brighter and fainter than 26.5, respectively.   
\label{fig12}}

\figcaption[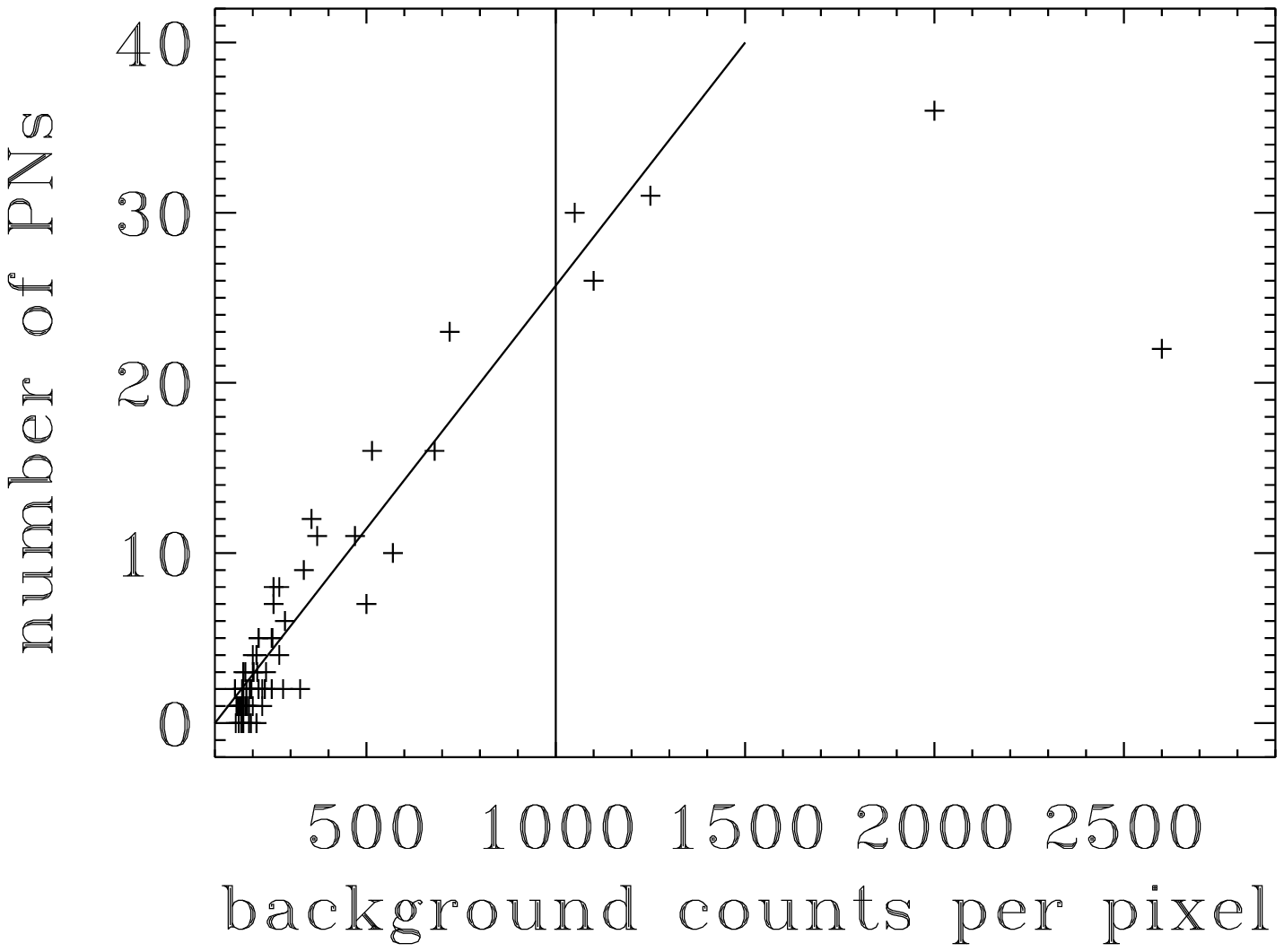]{The number of PNs brighter than $m$(5007)=27.6
detected within squares of 220$\times$220 pixels, plotted as a function
of the background counts per pixel in each square. The relation is linear
below 1000 counts per pixel, but not above that limit.
\label{fig13}}

\figcaption[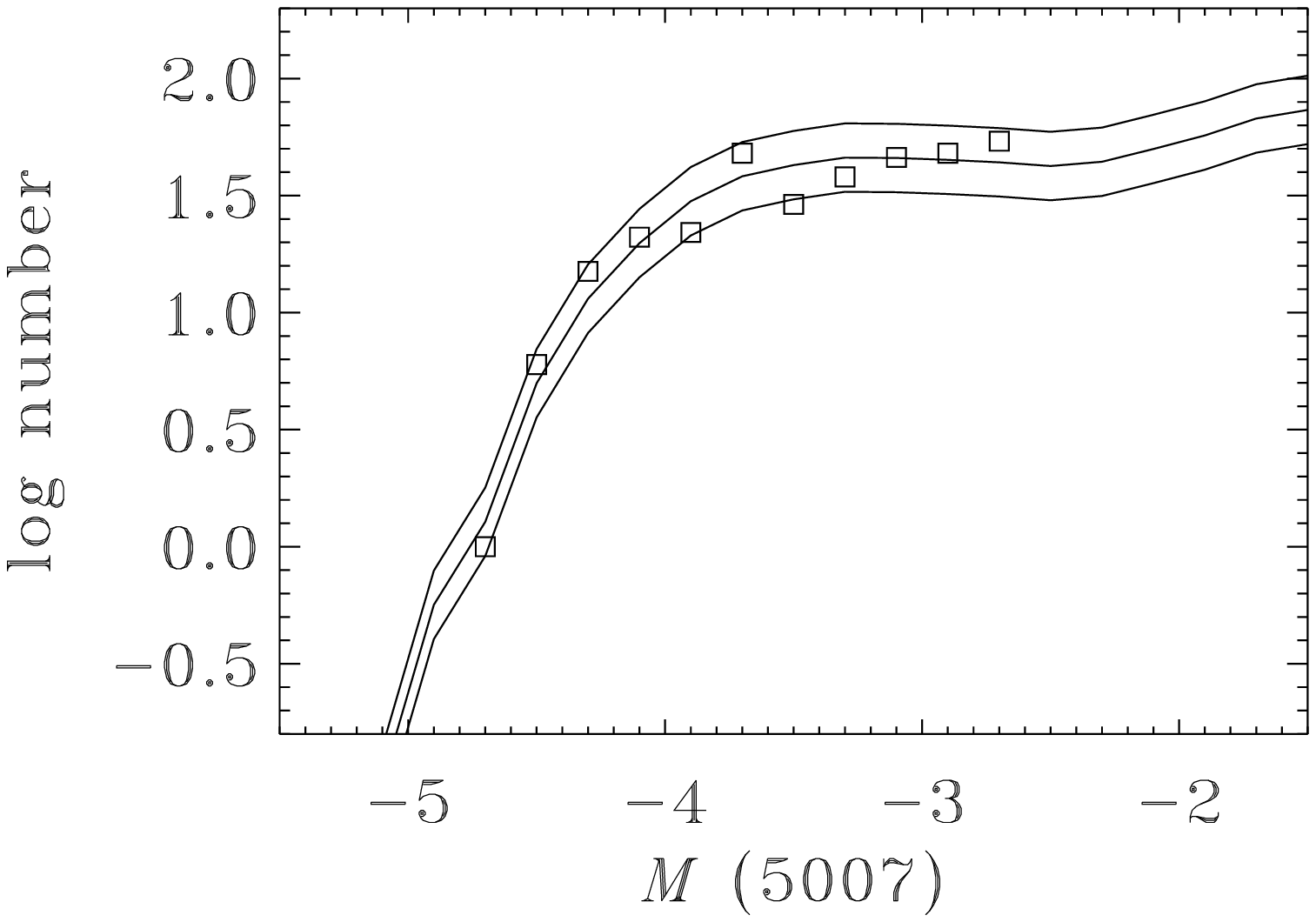]{The squares represent the observed [O {\sc III}]
$\lambda$5007 PNLF of NGC 4697, with the 328 data binned into 0.2 mag 
intervals. The apparent magnitudes $m$(5007) have been transformed into 
absolute magnitudes $M$(5007) by adopting 
an extinction correction of 0.105 mag 
and a distance modulus $m-M$ = 30.1. The three lines are PNLF simulations
(M\'endez and Soffner 1997) for three different sample sizes: 2500, 3500 
and 4900 PNs. From the sample size it is possible to estimate the PN 
formation rate (see text). The adoption of a distance modulus 29.9 or 
30.3 instead of 30.1 would ruin the fit.
\label{fig14}}

\figcaption[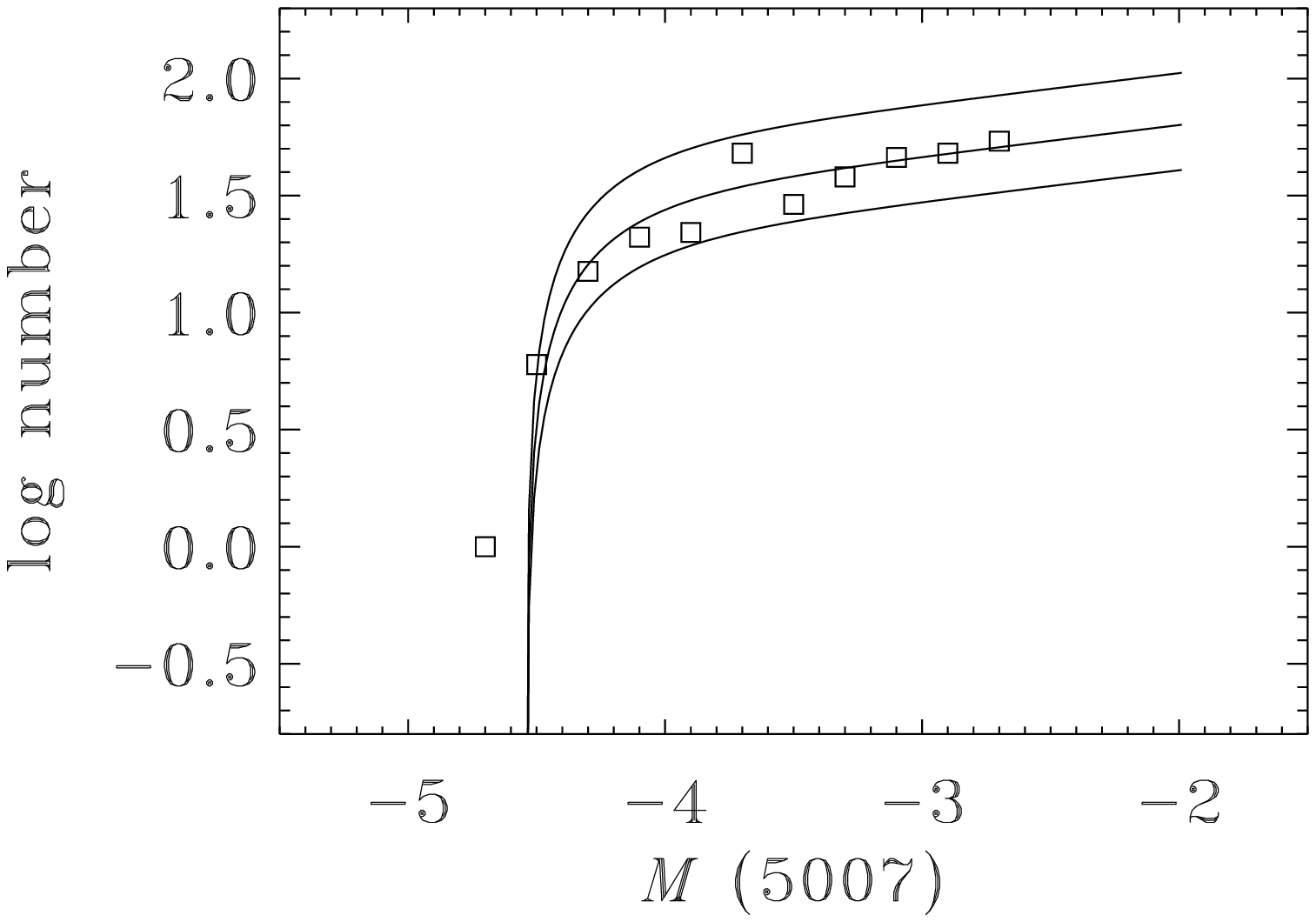]{The observed PNLF is now compared with the 
analytical representation (exponential + fixed or universal cutoff) 
proposed by Ciardullo et al. (1989). The adopted distance modulus is 
again 30.1. An attempt to fit the ``overluminous'' PN would result in 
a slightly smaller distance estimate. In fact the brightest PN is not 
``overluminous''; its existence is predicted by the simulations shown 
in Fig. 14. For sample sizes of the order of several thousands, as we 
have here, the ``universal cutoff'' representation of the PNLF begins 
to show some limitations, leading to slightly too small distances, as 
discussed by M\'endez (1999). 
\label{fig15}}

\figcaption[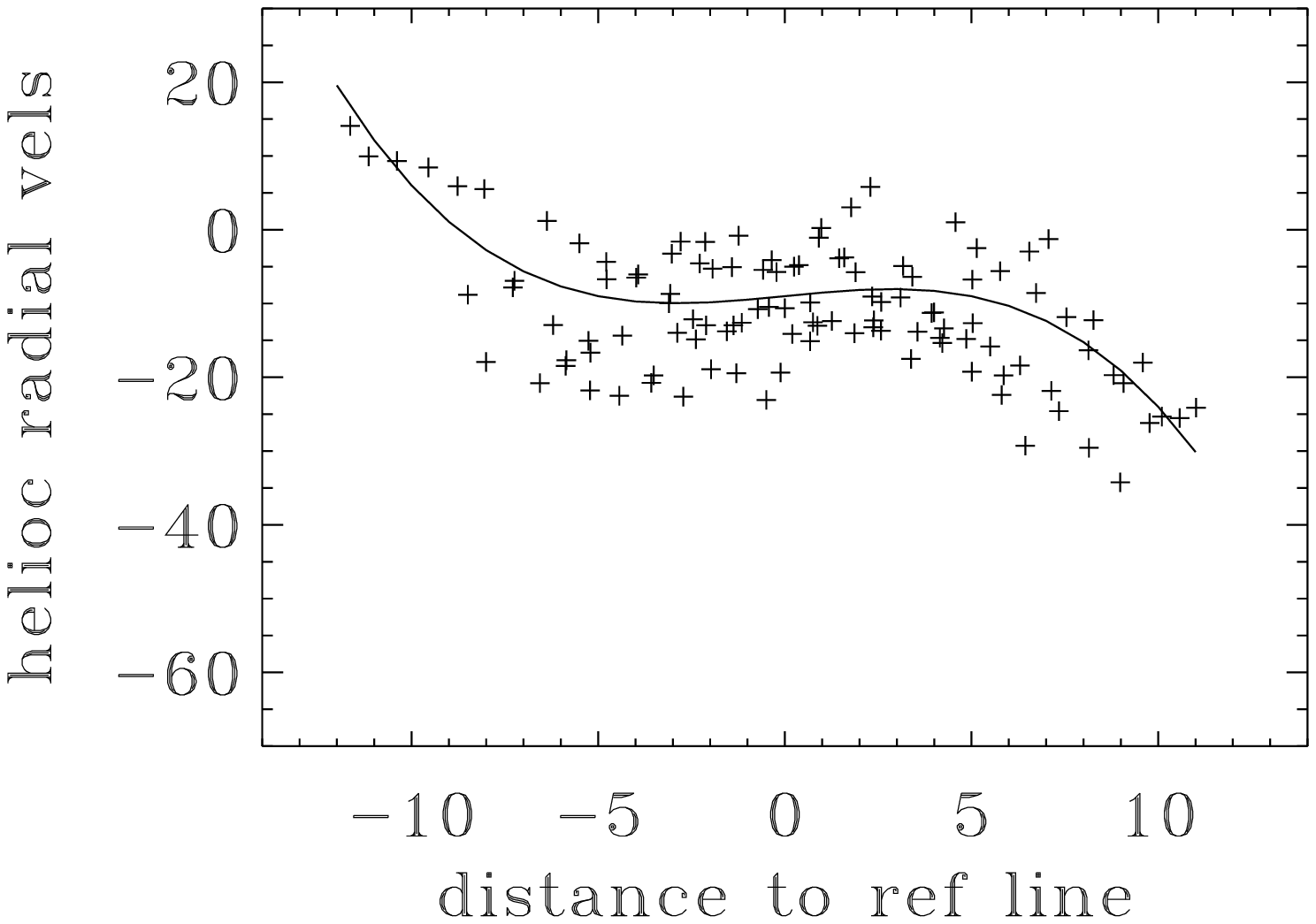]{The ``slitless'' heliocentric radial velocities 
of NGC 7293 measured at the 114 undispersed positions shown in Fig. 5. 
The velocities, in km s$^{-1}$, are plotted as a function of the 
distance from the undispersed position to the diagonal reference line 
shown in Fig. 5. These distances, defined positive above the reference 
line and negative below, are expressed in hundreds of pixels. The solid
line is a cubic parabola fitting the distribution.
\label{fig16}}

\figcaption[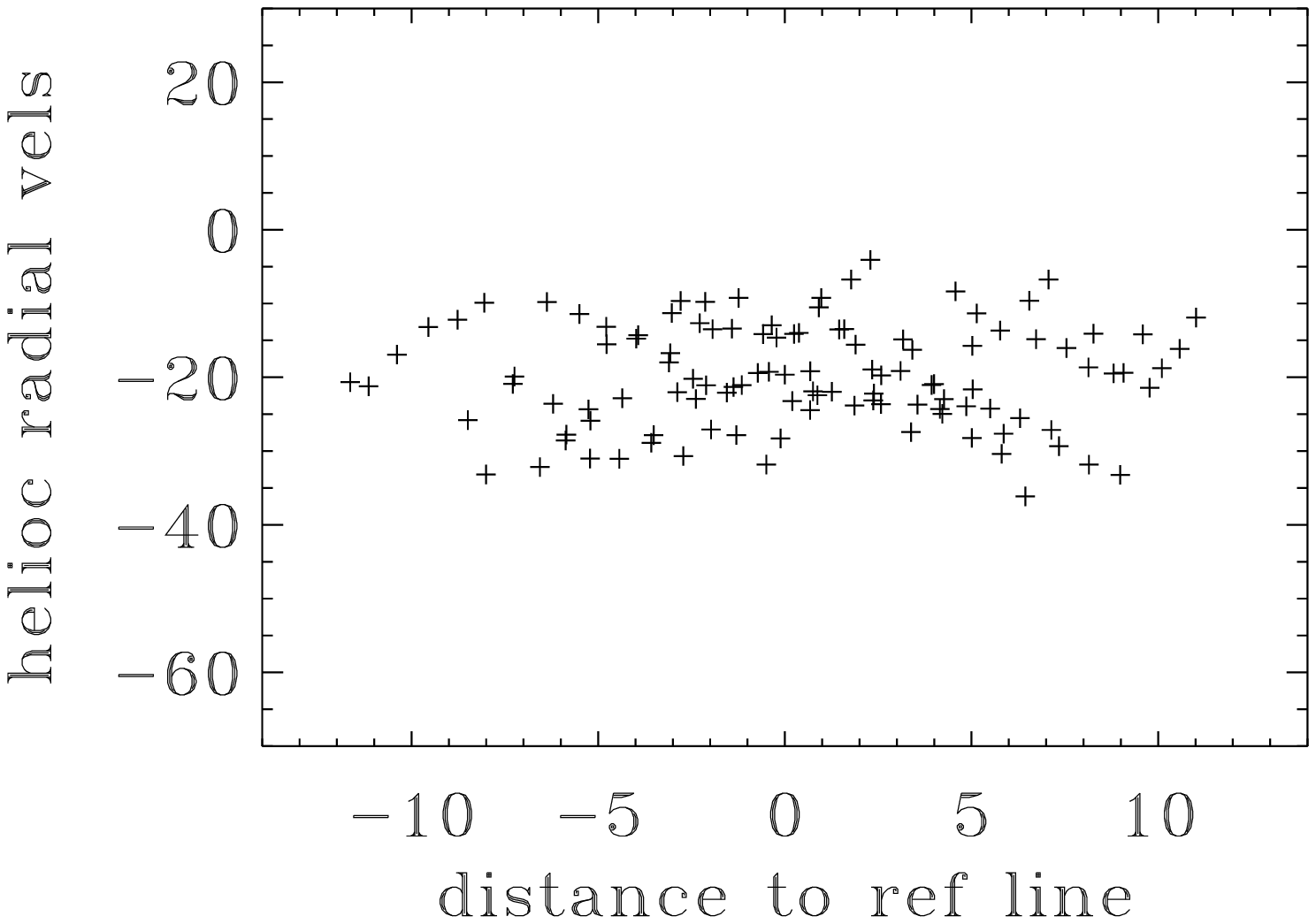]{The velocities of NGC 7293 plotted in Fig. 16 have 
been corrected as described in the text. From this figure we estimate that
the calibration errors in slitless velocities are below 20 km s$^{-1}$.
\label{fig17}}

\figcaption[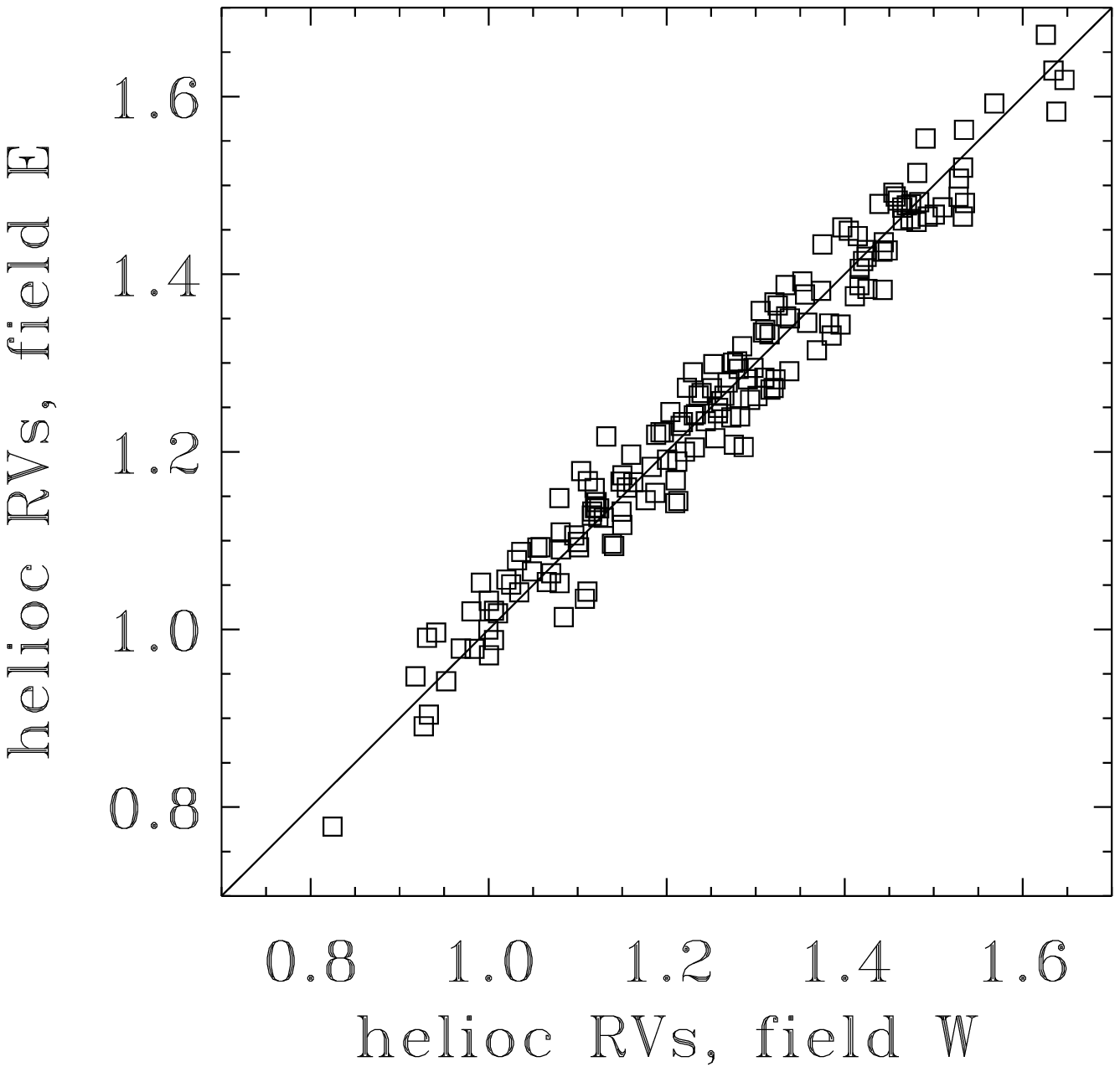]{The velocities of 165 PNs measured in both the E 
and W fields are compared. The velocities are expressed in thousands of 
km s$^{-1}$. The standard deviation is 36 km s$^{-1}$.
\label{fig18}}

\figcaption[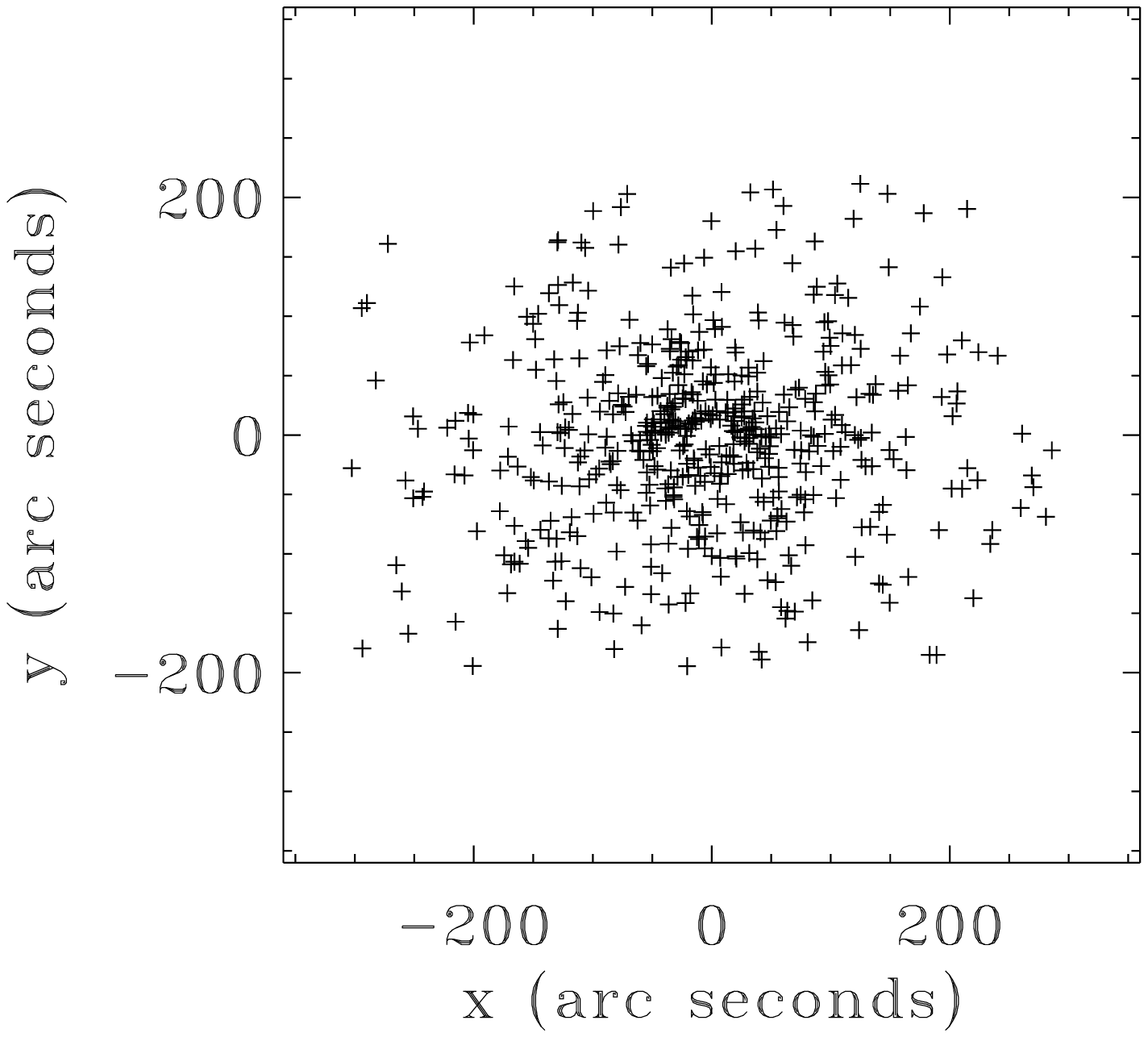]{Positions of the 535 PNs in arc seconds relative 
to the center of light of NGC 4697.
\label{fig19}}

\figcaption[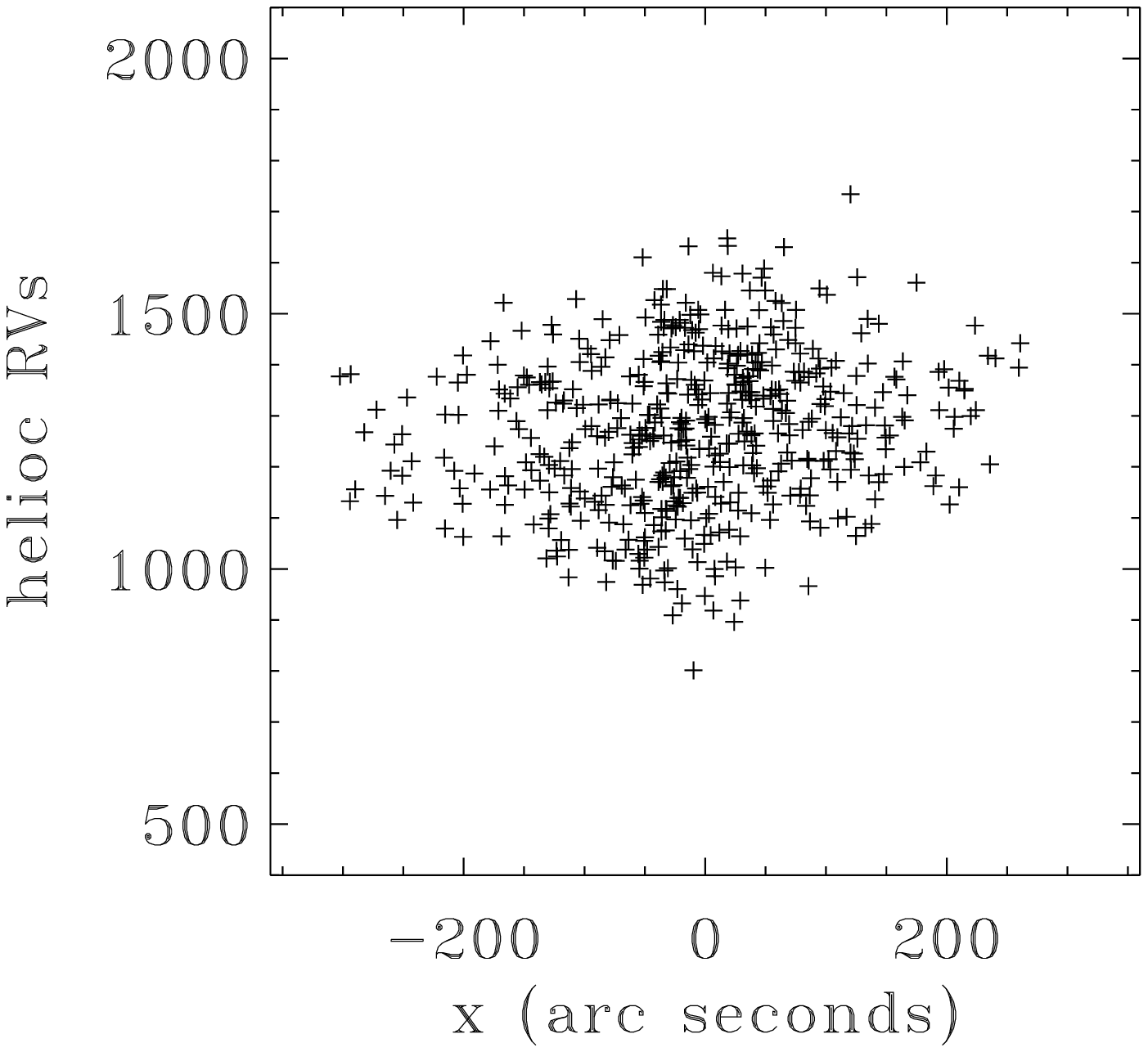]{Velocities of 531 PNs as a function of their $x$
coordinates in arc seconds relative to the center of light of NGC 4697.
\label{fig20}}

\figcaption[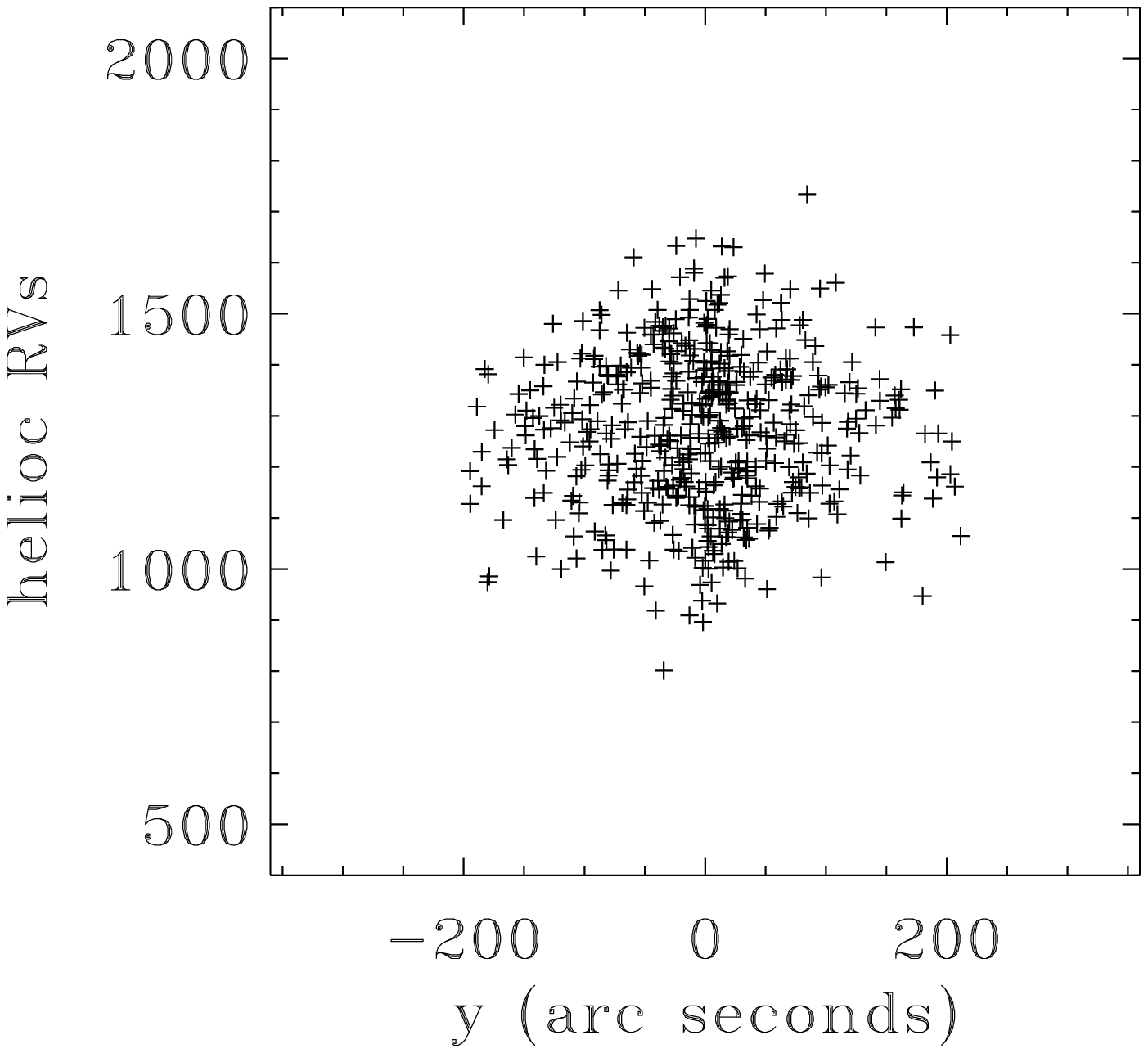]{Velocities of 531 PNs as a function of their $y$
coordinates in arc seconds relative to the center of light of NGC 4697.
\label{fig21}}

\figcaption[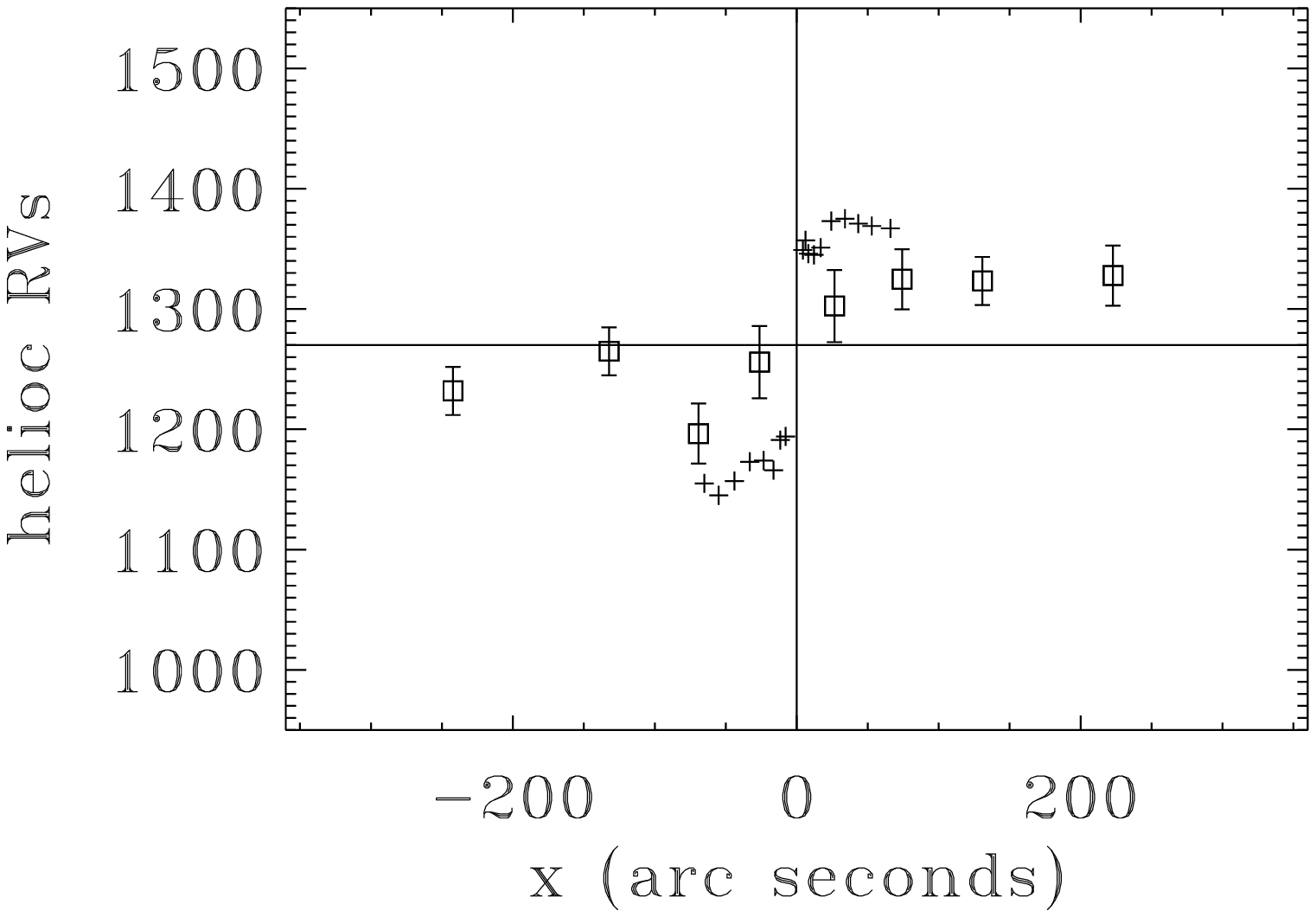]{The 531 PN velocities have been separated in 8 groups
according to their $x$ coordinates, and in the central groups (which include
many PNs) we have selected only PNs near the major axis. The $y$ 
coordinate limits for the 6 central groups, from left to right, are the 
following (in arc seconds): $\pm$40, $\pm$40, $\pm$20, $\pm$20, 
$\pm$40, $\pm$40. The squares with error bars indicate the average velocity 
within each group, plotted at the position of the average $x$ coordinate of 
the group. The numbers of PNs per group, from left to right, are the 
following: 26, 23, 34, 33, 39, 25, 23 and 16. The plus signs represent 
velocities measured on integrated light spectra along the major axis by 
Binney et al. (1990). The effective radius $R_{\rm e}$ is 95 arc seconds.
\label{fig22}}

\figcaption[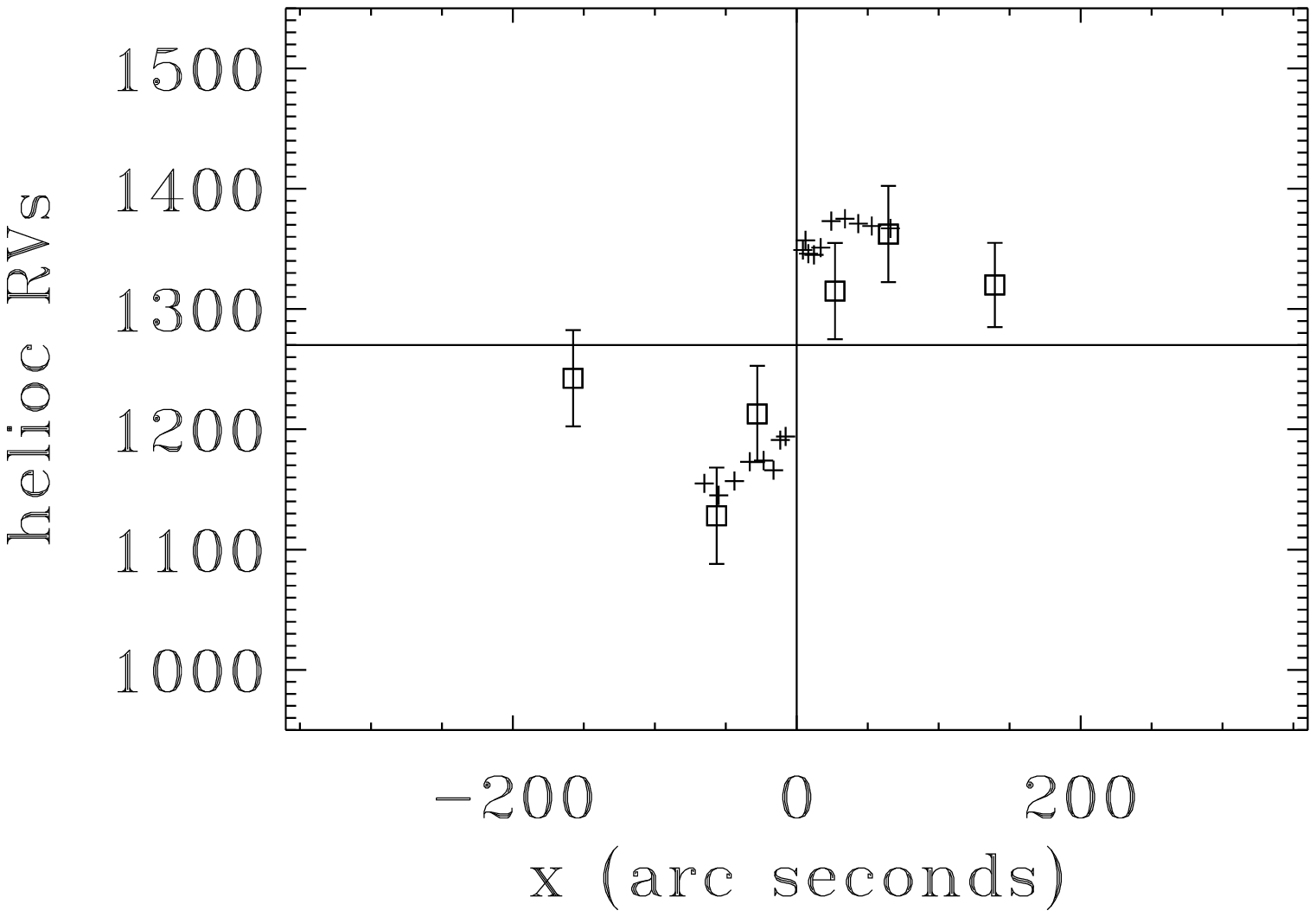]{Here we consider only velocities of PNs 
within $\pm$10 arc seconds of the major axis. 
The selected PN velocities have been separated in 6 groups
according to their $x$ coordinates. The squares with error bars
indicate the average velocity within each group, plotted at the 
position of the average $x$ coordinate of the group. The numbers 
of PNs per group, from left to right, are the following: 11, 14, 
15, 16, 12, and 9. The plus signs have the same meaning as in Fig. 22.
\label{fig23}}

\figcaption[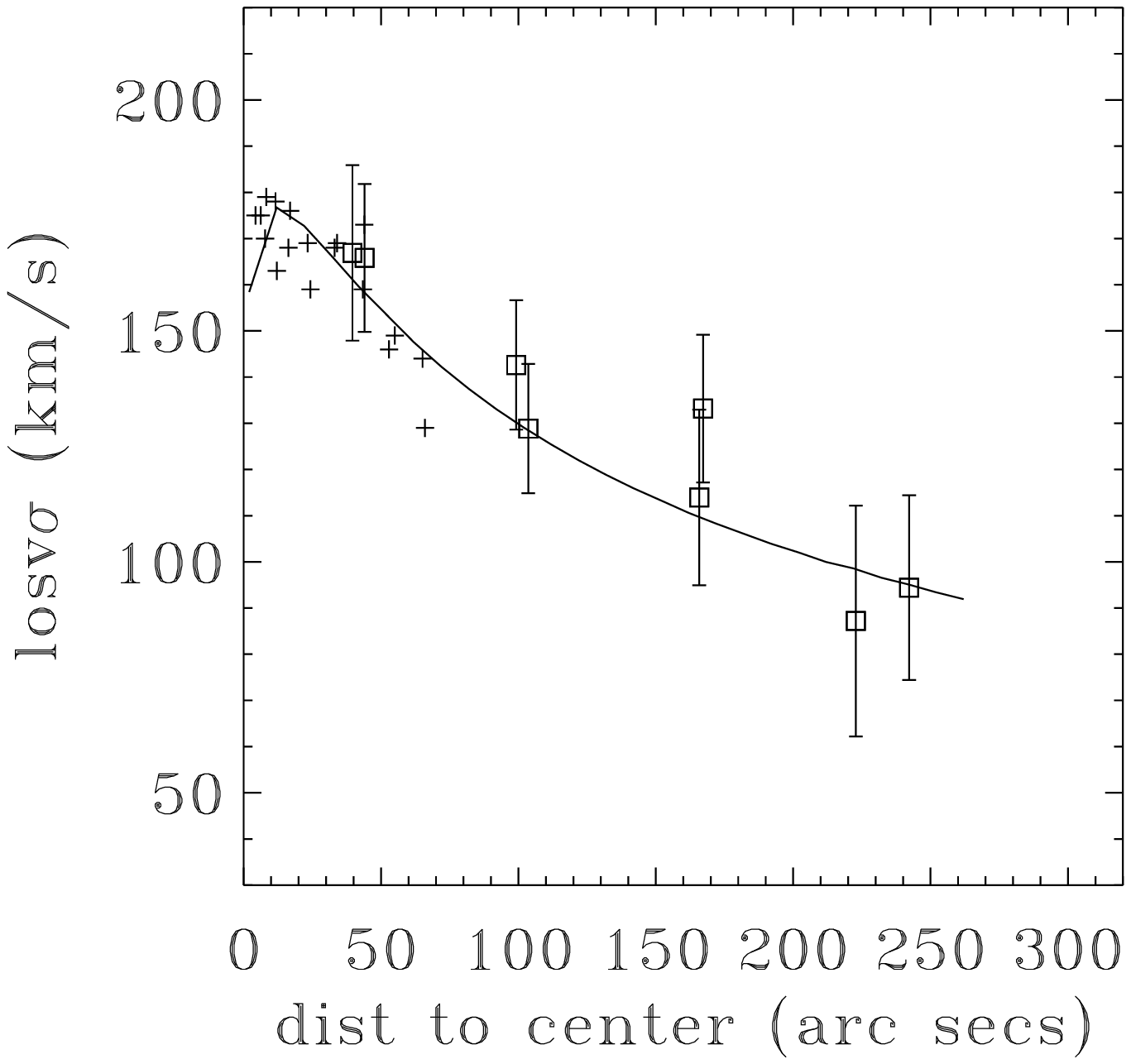]{The plus signs are line-of-sight velocity 
dispersions measured by Binney et al. (1990) on integrated light 
spectra along the major axis. The squares with error bars are our 
line-of-sight velocity dispersions from the PN velocities. The PNs 
have been separated in 8 groups according to angular 
distance from the center, as explained in the text. 
The numbers of PNs per group are: center NE, 108; center SW, 83; 
inner ring NE, 81; inner ring SW, 92; second ring NE, 57; 
second ring SW, 39; outer zone NE, 26; outer zone SW, 16. 
The solid line is the analytical model by Hernquist (1990), with a 
constant M/L ratio and a total mass of 1.9$\times 10^{11}$ solar masses, 
adopting $R_{\rm e}$ = 95 arc seconds. A smaller $R_{\rm e}$ of 75 arc 
seconds would give a 10\% smaller total mass.
\label{fig24}}

\newpage

\begin{figure}
\figurenum{1}
\epsscale{1.0}
\plotone{f1.ps}
\caption{Figure 1}
\end{figure}

\begin{figure}
\figurenum{2}
\epsscale{1.0}
\plotone{f2.ps}
\caption{Figure 2}
\end{figure}

\begin{figure}
\figurenum{3}
\epsscale{1.0}
\plotone{f3.eps}
\caption{Figure 3}
\end{figure}

\begin{figure}
\figurenum{4}
\epsscale{1.0}
\plotone{f4.eps}
\caption{Figure 4}
\end{figure}

\begin{figure}
\figurenum{5}
\epsscale{1.0}
\plotone{f5.ps}
\caption{Figure 5}
\end{figure}

\begin{figure}
\figurenum{6}
\epsscale{1.0}
\plotone{f6.eps}
\caption{Figure 6}
\end{figure}

\begin{figure}
\figurenum{7}
\epsscale{1.0}
\plotone{f7.eps}
\caption{Figure 7}
\end{figure}

\begin{figure}
\figurenum{8}
\epsscale{1.0}
\plotone{f8.eps}
\caption{Figure 8}
\end{figure}

\begin{figure}
\figurenum{9}
\epsscale{1.0}
\plotone{f9.eps}
\caption{Figure 9}
\end{figure}

\begin{figure}
\figurenum{10}
\epsscale{1.0}
\plotone{f10.eps}
\caption{Figure 10}
\end{figure}

\begin{figure}
\figurenum{11}
\epsscale{1.0}
\plotone{f11.eps}
\caption{Figure 11}
\end{figure}

\begin{figure}
\figurenum{12}
\epsscale{1.0}
\plotone{f12.ps}
\caption{Figure 12}
\end{figure}

\begin{figure}
\figurenum{13}
\epsscale{1.0}
\plotone{f13.ps}
\caption{Figure 13}
\end{figure}

\begin{figure}
\figurenum{14}
\epsscale{1.0}
\plotone{f14.ps}
\caption{Figure 14}
\end{figure}

\begin{figure}
\figurenum{15}
\epsscale{1.0}
\plotone{f15.ps}
\caption{Figure 15}
\end{figure}

\begin{figure}
\figurenum{16}
\epsscale{1.0}
\plotone{f16.ps}
\caption{Figure 16}
\end{figure}

\begin{figure}
\figurenum{17}
\epsscale{1.0}
\plotone{f17.ps}
\caption{Figure 17}
\end{figure}

\begin{figure}
\figurenum{18}
\epsscale{1.0}
\plotone{f18.ps}
\caption{Figure 18}
\end{figure}

\clearpage

\begin{figure}
\figurenum{19}
\epsscale{1.0}
\plotone{f19.ps}
\caption{Figure 19}
\end{figure}

\begin{figure}
\figurenum{20}
\epsscale{1.0}
\plotone{f20.ps}
\caption{Figure 20}
\end{figure}

\begin{figure}
\figurenum{21}
\epsscale{1.0}
\plotone{f21.ps}
\caption{Figure 21}
\end{figure}

\begin{figure}
\figurenum{22}
\epsscale{1.0}
\plotone{f22.ps}
\caption{Figure 22}
\end{figure}

\begin{figure}
\figurenum{23}
\epsscale{1.0}
\plotone{f23.ps}
\caption{Figure 23}
\end{figure}

\begin{figure}
\figurenum{24}
\epsscale{1.0}
\plotone{f24.ps}
\caption{Figure 24}
\end{figure}

\begin{deluxetable}{lrlrc}
\tablecaption{Observations and calibrations \label{tbl-1}}
\tablewidth{0pt}
\tablehead{
\colhead{Field} & \colhead{Config} &
\colhead{FORS1 CCD frame identification} &
\colhead{exp (s)} & \colhead{Air mass\tablenotemark{a}}}
\startdata
 He 2-118    & on-band  & FORS.1999-04-20T08:36:34.818.fits &   20 & 1.26 \\
 He 2-118    & grism+on & FORS.1999-04-20T08:37:56.133.fits &   20 & 1.26 \\
 NGC 4697 W  & off-band & FORS.1999-04-22T01:40:24.000.fits &  600 & 1.17 \\
 NGC 4697 W  & on-band  & FORS.1999-04-22T01:51:23.924.fits & 1200 & 1.14 \\
 NGC 4697 W  & grism+on & FORS.1999-04-22T02:12:35.052.fits & 2400 & 1.10 \\
 NGC 4697 W  & off-band & FORS.1999-04-22T03:01:23.624.fits &  900 & 1.06 \\
 NGC 4697 W  & on-band  & FORS.1999-04-22T03:17:23.062.fits & 1200 & 1.06 \\
 NGC 4697 W  & grism+on & FORS.1999-04-22T03:38:32.814.fits & 2400 & 1.07 \\
 NGC 4697 W  & grism+on & FORS.1999-04-22T04:24:25.502.fits & 2400 & 1.11 \\
 NGC 4697 W  & off-band & FORS.1999-04-22T05:14:19.274.fits &  900 & 1.19 \\
 NGC 4697 W  & on-band  & FORS.1999-04-22T05:30:18.150.fits & 1200 & 1.24 \\
 He 2-118    & on-band  & FORS.1999-04-22T08:25:39.484.fits &   10 & 1.27 \\
 He 2-118    & grism+on & FORS.1999-04-22T08:26:56.090.fits &   20 & 1.27 \\
 NGC 4697 E  & off-band & FORS.2000-05-29T23:09:36.121.fits &  900 & 1.16 \\
 NGC 4697 E  & on-band  & FORS.2000-05-29T23:25:48.288.fits & 1500 & 1.12 \\
 NGC 4697 E  & grism+on & FORS.2000-05-29T23:52:02.001.fits & 2400 & 1.08 \\
 NGC 4697 E  & off-band & FORS.2000-05-30T00:35:04.237.fits &  900 & 1.06 \\
 NGC 4697 E  & on-band  & FORS.2000-05-30T00:51:11.048.fits & 1500 & 1.06 \\
 NGC 4697 E  & grism+on & FORS.2000-05-30T01:17:23.306.fits & 2400 & 1.08 \\
 NGC 4697 E  & off-band & FORS.2000-05-30T01:59:12.334.fits &  900 & 1.10 \\
 NGC 4697 E  & on-band  & FORS.2000-05-30T02:15:15.528.fits & 1500 & 1.14 \\
 NGC 4697 E  & grism+on & FORS.2000-05-30T02:41:28.870.fits & 2400 & 1.23 \\
 G 138-31    & on-band  & FORS.2000-05-30T04:07:08.131.fits &  100 & 1.21 \\
 NGC 7293 p1 & on-band  & FORS.2000-05-30T09:33:18.900.fits &  100 & 1.04 \\
 NGC 7293 p1 & grism+on & FORS.2000-05-30T09:36:30.313.fits &  250 & 1.03 \\
 NGC 7293 p2 & on-band  & FORS.2000-05-30T10:23:51.849.fits &  100 & 1.00 \\
 NGC 7293 p2 & grism+on & FORS.2000-05-30T10:27:02.299.fits &  250 & 1.00 \\
 MOS cal. p1 & undisp.  & FORS.2000-05-30T11:37:59.703.fits &   10 & $-$  \\
 MOS cal. p1 & disp.    & FORS.2000-05-30T11:40:03.923.fits &  300 & $-$  \\ 
 MOS cal. p2 & undisp.  & FORS.2000-05-30T11:54:53.993.fits &   10 & $-$  \\
 MOS cal. p2 & disp.    & FORS.2000-05-30T11:56:58.370.fits &  300 & $-$  \\
 MOS cal. p3 & undisp.  & FORS.2000-05-30T12:29:15.781.fits &   10 & $-$  \\
 MOS cal. p3 & disp.    & FORS.2000-05-30T12:31:20.217.fits &  300 & $-$  \\
 MOS cal. p4 & undisp.  & FORS.2000-05-30T12:48:53.658.fits &   10 & $-$  \\
 MOS cal. p4 & disp.    & FORS.2000-05-30T12:50:58.267.fits &  300 & $-$  \\
 MOS cal. p5 & undisp.  & FORS.2000-05-30T13:07:27.466.fits &   10 & $-$  \\
 MOS cal. p5 & disp.    & FORS.2000-05-30T13:09:32.060.fits &  300 & $-$  \\
 MOS cal. p6 & undisp.  & FORS.2000-05-30T13:33:04.556.fits &   10 & $-$  \\
 MOS cal. p6 & disp.    & FORS.2000-05-30T13:35:09.123.fits &  300 & $-$  \\
 MOS cal. p7 & undisp.  & FORS.2000-05-30T13:57:20.765.fits &   10 & $-$  \\
 MOS cal. p7 & disp.    & FORS.2000-05-30T13:59:25.251.fits &  300 & $-$  \\
 MOS cal. p8 & undisp.  & FORS.2000-05-30T14:17:05.835.fits &   10 & $-$  \\
 MOS cal. p8 & disp.    & FORS.2000-05-30T14:19:10.277.fits &  300 & $-$  \\
 MOS cal. p9 & undisp.  & FORS.2000-05-30T14:49:00.448.fits &   10 & $-$  \\
 MOS cal. p9 & disp.    & FORS.2000-05-30T14:51:05.629.fits &  300 & $-$  \\
 MOS cal. p10 & undisp. & FORS.2000-05-30T15:13:21.145.fits &   10 & $-$  \\
 MOS cal. p10 & disp.   & FORS.2000-05-30T15:15:25.757.fits &  300 & $-$  \\
 NGC 4697 W  & off-band & FORS.2000-05-30T23:01:16.768.fits &  900 & 1.17 \\
 NGC 4697 W  & off-band & FORS.2000-05-30T23:17:22.665.fits &  900 & 1.14 \\
 NGC 4697 W  & on-band  & FORS.2000-05-30T23:33:19.758.fits & 1500 & 1.11 \\
 NGC 4697 W  & grism+on & FORS.2000-05-31T00:00:04.556.fits & 2400 & 1.07 \\
 NGC 4697 W  & on-band  & FORS.2000-05-31T00:42:06.004.fits & 1800 & 1.06 \\
 NGC 4697 E  & off-band & FORS.2000-05-31T01:56:56.533.fits &  900 & 1.10 \\
 NGC 4697 E  & on-band  & FORS.2000-05-31T02:12:50.309.fits & 1500 & 1.14 \\
 NGC 4697 E  & grism+on & FORS.2000-05-31T02:38:59.384.fits & 2400 & 1.23 \\
 G 138-31    & on-band  & FORS.2000-05-31T03:24:08.942.fits &  100 & 1.26 \\
 NGC 7293 p3 & on-band  & FORS.2000-05-31T08:34:14.790.fits &  100 & 1.13 \\
 NGC 7293 p3 & grism+on & FORS.2000-05-31T08:37:05.265.fits &  250 & 1.12 \\
 NGC 7293 p4 & on-band  & FORS.2000-05-31T08:54:43.811.fits &  100 & 1.09 \\
 NGC 7293 p4 & grism+on & FORS.2000-05-31T08:57:33.414.fits &  250 & 1.08 \\
 NGC 7293 p5 & on-band  & FORS.2000-05-31T09:13:37.789.fits &  100 & 1.06 \\
 NGC 7293 p5 & grism+on & FORS.2000-05-31T09:16:28.264.fits &  250 & 1.05 \\
 NGC 7293 p6 & on-band  & FORS.2000-05-31T09:58:28.841.fits &  100 & 1.01 \\
 NGC 7293 p6 & grism+on & FORS.2000-05-31T10:01:18.615.fits &  250 & 1.01 \\
 NGC 4697 E  & off-band & FORS.2000-05-31T23:15:10.254.fits &  900 & 1.14 \\
 NGC 4697 E  & on-band  & FORS.2000-05-31T23:31:04.416.fits & 1500 & 1.10 \\
 NGC 4697 E  & grism+on & FORS.2000-05-31T23:57:14.651.fits & 2400 & 1.07 \\
 NGC 4697 W  & off-band & FORS.2000-06-01T00:43:25.603.fits &  900 & 1.06 \\
 NGC 4697 W  & on-band  & FORS.2000-06-01T00:59:19.970.fits & 1500 & 1.06 \\
 NGC 4697 W  & grism+on & FORS.2000-06-01T01:25:29.402.fits & 2400 & 1.09 \\
\enddata
\tablenotetext{a}{the air masses correspond to the middle of each exposure}
\end{deluxetable}

\end{document}